\documentclass[structabstract]{aa}
\usepackage{graphicx}
\usepackage{longtable,lscape}
\usepackage[authoryear]{natbib}
\bibpunct{(}{)}{;}{a}{}{,}
\usepackage{txfonts}
\begin{document}
\title{The Double-Peaked 2008 Outburst of the\\ Accreting Milli-Second X-ray Pulsar,\\ IGR J00291+5934} 
\titlerunning{The 2008 Outburst of IGR J00291+5934}

\author{F.~Lewis\inst{1,2,3}
  \and D.M.~Russell\inst{4}
  \and P.G.~Jonker\inst{5,6,7}
  \and M.~Linares\inst{4,8}
  \and V.~Tudose\inst{9,10,11}
  \and P.~Roche\inst{1,2,3} 
  \and J.S.~Clark\inst{2}
  \and M.A.P.~Torres\inst{6}
  \and D.~Maitra\inst{4,12}
  \and C.G.~Bassa\inst{13}
  \and D.~Steeghs\inst{6,14}
  \and A.~Patruno\inst{4}
  \and S.~Migliari\inst{15}
  \and R.~Wijnands\inst{4}
  \and G.~Nelemans\inst{7}
  \and L.J.~Kewley\inst{16}
  \and V.E.~Stroud\inst{1,2,3}
  \and M.~Modjaz\inst{17,18}
  \and J.S.~Bloom\inst{17}
  \and C.H.~Blake\inst{6}
  \and D.~Starr\inst{17,19}
}

\offprints{Fraser Lewis, \email{fraser.lewis@faulkes-telescope.com}}

\institute{Faulkes Telescope Project, School of Physics and Astronomy, Cardiff University, 5, The Parade, Cardiff, CF24 3AA, Wales
 \and Department of Physics and Astronomy, The Open University, Walton Hall, Milton Keynes, MK7 6AA, England 
 \and Division of Earth, Space and Environment, University of Glamorgan, Pontypridd, CF37 1DL, Wales
 \and Astronomical Institute `Anton Pannekoek', University of Amsterdam, P.O. Box 94249, 1090 GE Amsterdam, the Netherlands
 \and SRON, Netherlands Institute for Space Research, Sorbonnelaan 2, 3584 CA, Utrecht, the Netherlands 
 \and Harvard-Smithsonian Center for Astrophysics, 60 Garden Street, Cambridge, MA 02138, USA 
 \and Department of Astrophysics, IMAPP, Radboud University Nijmegen, Toernooiveld 1, 6525 ED, Nijmegen, the Netherlands
 \and MIT Kavli Institute for Astrophysics and Space Research, 70 Vassar Street, Cambridge, MA 02139,USA
 \and Netherlands Institute for Radio Astronomy, Oude Hoogeveensedijk 4, 7991 PD Dwingeloo, the Netherlands
 \and Astronomical Institute of the Romanian Academy, Cutitul de Argint 5, RO-040557 Bucharest, Romania
 \and Research Center for Atomic Physics and Astrophysics, Atomistilor 405, RO-077125 Bucharest, Romania
 \and Department of Astronomy, University of Michigan, 500 Church Street, Ann Arbor, MI 48109, USA 
 \and Jodrell Bank Centre for Astrophysics, School of Physics and Astronomy, University of Manchester, Manchester M13 9PL, England
 \and Department of Physics, University of Warwick, Coventry, CV4 7AL, England 
 \and ESAC, Urb. Villafranca del Castillo, P.O. Box 50727, 28080 Madrid, Spain
 \and University of Hawaii, 2680 Woodlawn Drive, Honolulu, HI 96822, USA
 \and Department of Astronomy, University of California, Berkeley, 601 Campbell Hall, Berkeley, CA 94720, USA
 \and Miller Fellow
 \and Las Cumbres Observatory Global Telescope Project, 6740 Cortona Dr., Santa Barbara, CA 93117, USA
}

\date{Received / Accepted}

 \abstract 
{In August 2008, the accreting milli-second X-ray pulsar (AMXP), IGR J00291+5934, underwent an outburst lasting $\sim$ 100 days, the first since its discovery in 2004.}
{We present data from the 2008 double-peaked outburst of IGR J00291+5934 from Faulkes Telescope North, the Isaac Newton Telescope, the Keck Telescope, PAIRITEL, the Westerbork Synthesis Radio Telescope and the Swift, XMM-Newton and RXTE X-ray missions. We study the outburst's evolution at various wavelengths, allowing us to probe accretion physics in this AMXP.}
{We study the light curve morphology, presenting the first radio--X--ray Spectral Energy Distributions (SEDs) for this source and the most detailed UV--IR SEDs for any outbursting AMXP. We show simple models that attempt to identify the emission mechanisms responsible for the SEDs. We analyse short-timescale optical variability, and compare a medium resolution optical spectrum with those from 2004.} 
{The outburst morphology is unusual for an AMXP, comprising two peaks, the second containing a `plateau' of $\sim$ 10 days at maximum brightness within 30 days of the initial activity. This has implications on duty cycles of short-period X-ray transients. The X-ray spectrum can be fitted by a single, hard power-law. We detect optical variability of $\sim$ 0.05 magnitudes, on timescales of minutes, but find no periodic modulation. In the optical, the SEDs contain a blue component, indicative of an irradiated disc, and a transient near-infrared (NIR) excess. This excess is consistent with a simple model of an optically thick synchrotron jet (as seen in other outbursting AMXPs), however we discuss other potential origins. The optical spectrum shows a double-peaked H$\alpha$ profile, a diagnostic of an accretion disc, but we do not clearly see other lines (e.g. He I, II) that were reported in 2004.} 
{Optical/IR observations of AMXPs appear to be excellent for studying the evolution of both the outer accretion disc and the inner jet, and may eventually provide us with tight constraints to model disc-jet coupling in accreting neutron stars.} 

\maketitle   

%

\section{Introduction} 
\subsection{Accreting Milli-Second X-ray Pulsars} 

Low-Mass X-ray Binaries (LMXBs) comprise a compact primary (neutron star or black hole) and a low-mass (usually K or M class) secondary with the compact object accreting material from its companion by means of Roche lobe overflow. It was postulated in the early 1980s that accreting milli-second X-ray pulsars would form a sub group of neutron star LMXBs (e.g. \citealp{rad,alp}) and thus prove to be the link between accreting LMXBs and isolated milli-second pulsars. The mechanism believed to be responsible is the spin-up of an old, weakly magnetized, neutron star caused by the accretion of matter and angular momentum from a donor star. This would result in a compact transient X-ray binary (its transient nature demonstrating a period of increased mass transfer rate onto the neutron star) with the neutron star spinning so rapidly that, once the accretion has ceased, it can once again become a rapid radio pulsar (see \citealp{wij06} and references therein). This was subsequently confirmed observationally with the discovery of pulsations from the transient source SAX J1808.4$-$3658 \citep{wij98} in 1998 April and the first LMXB to be detected as a milli-second radio pulsar, FIRST J102347.6+003841 \citep{arc}. 

Properties of AMXPs include weaker peak luminosities in outburst than `classical' (i.e. not showing pulsations during outburst) neutron star LMXBs, extremely low-mass companions and faint quiescent X-ray luminosities ($\le$ 10$^{32}$ erg s$^{-1}$), dominated by a hard power-law component (see \citealp{wij06,wij} and references therein). For most AMXPs, the duty cycle is a few weeks of outburst every few years. Taking into account the recent (2009 August and September respectively) discoveries of NGC 6440 X-2 \citep{alt,hei} and IGR J17511$-$305 \citep{mar09,boz}, there are currently twelve known AMXPs \citep{wij}. Observationally, these systems have orbital periods of $P_\mathrm{orb}$ $\sim$ 40 minutes - 4 hours. Little is known about the optical counterparts of these systems; many have not been detected in the optical/infrared regime, even in outburst (e.g. XTE J1751-305; \citealp{jon03}). Some LMXBs suffer from a high level of absorption from interstellar dust and/or lie at large distances, and those systems with detections, particularly AMXPs, are often very faint (due to their relatively low-mass donors), especially when in quiescence (see \citealp{dav09}).

\subsection{The Accreting Milli-Second X-ray Pulsar IGR J00291+5934}

The accreting millisecond X-ray pulsar IGR J00291+5934 (hereafter ``00291+59'') was first discovered on 2004 December 2 (MJD 53341) in outburst during a routine Galactic Plane Scan of the Cassiopeia region by INTEGRAL \citep{eck}. It was given the co-ordinates RA 00:29.1, dec +59:34 (J2000.0) with an error radius of 1\farcm5 (l = 120\fdg1, b = $-$3\fdg2). In three successive pointings, the IBIS/ISGRI instrument recorded an average flux of 55 $\pm$ 5 mCrab (20 -- 60 keV). During its only pointing, JEM-X detected the source with a flux of 23 $\pm$ 5 mCrab (3 -- 10 keV) \citep{eck}. On the following day, using the RXTE PCA (Rossi X-ray Timing Explorer Proportional Counter Array), 00291+59 was confirmed as an X-ray pulsar with a spin of 598.88 Hz \citep{mark04a}, corresponding to a spin period of $\sim$  1.7 ms. This made it the fastest known AMXP (and the 6th one to be discovered). A sinusoidal frequency modulation of 147.4 minutes (2.46 hours) found in X-ray (by RXTE) was interpreted as the orbital period and no evidence of X-ray eclipses was present \citep{mark04b}. The fast X-ray variability was stronger and had lower characteristic frequencies than any other neutron star LMXB, being more similar to that of black hole systems \citep{lin07}.

Within days, optical and infrared counterparts were identified  at RA 00:29:03.06, dec +59:34:19.0 (J2000.0) to an uncertainty of $\sim$ 0\farcs5 \citep{fox,ste}. The source was bright with magnitudes of R $\sim$ 17.4; J = 16.8 $\pm$ 0.1; H = 16.8 $\pm$ 0.3; K = 16.1 $\pm$ 0.2. Later observations during quiescence gave values of R = 23.1 $\pm$ 0.1 in 2005 October and K = 19.0 $\pm$ 0.1 in 2005 January \citep{tor08a}. During its decay, variability of $\sim$ 0.3 magnitudes (not associated with the system's orbital period) was noted on timescales of seconds to hours in R-band \citep{bik}. An optical spectrum of 00291+59 was taken on December 5 by the William Herschel Telescope (WHT) showing `weak evidence' for broad emission features at H$\alpha$ and He \small II\normalsize\ (4686\ \AA) \citep{roe,tor08a}, both of which are considered diagnostics for outbursting LMXBs. A further spectrum was taken on December 12 (still during outburst) using the 10-metre Keck 1 telescope \citep{fil,rey} which  detected broad (FWHM = 1200 km s$^{-1}$) emission at H$\alpha$, H$\beta$ and He \small I\normalsize\ (6678\ \AA) as well as narrow (FWHM = 300 km s$^{-1}$) `very weak' features at He \small II\normalsize\ (4686\ \AA). The optical-IR SED during outburst was dominated by thermal emission with a NIR excess thought to be due to synchrotron emission \citep{tor08a}. Until the 2008 outburst, no further spectra of 00291+59 had been published.

On 2004 December 4, radio observations with the Ryle Telescope, Cambridge yielded a `probable detection' of 1.1 mJy at 15 GHz \citep{pool}. 12 hours later, at the same frequency, this signal had disappeared below the detection threshold of $\sim$ 0.6 mJy \citep{fen04a}, suggesting that this was indeed the transient source. A day later, the Westerbork Synthesis Radio Telescope (WSRT) detected a radio source with mean flux density of 0.250 $\pm$ 0.035 mJy (at 5 GHz), with evidence of this fading during the 10 hour observation \citep{fen04a}. Very Large Array (VLA) observations on December 9 gave a detection of 0.17 $\pm$ 0.05 mJy at 4.86 GHz \citep{rup}.

A reliable distance to the system is still to be determined. An upper limit of 3.3 kpc based on X-ray absorption has been suggested \citep{sha}, however \citet{bur} suggest a value of between 7.4 and 10.7 kpc, based upon the pulsar's spin-down rate. Most recently, an estimate of 2 -- 4 kpc has been made based upon the X-ray luminosity during outburst \citep{tor08a}. An upper limit on the system's donor has been estimated at 0.16 M$_{\sun}$, implying it is most likely a hot brown dwarf \citep{gall}. This was calculated from its mass function and also its similarities with the best studied AMXP, SAX J1808.4$-$3658. 

Observations of 00291+59 were taken in quiescence with the 3.6 metre Italian TNG telescope. The data were taken over 3 nights in 2005 August (V and R-bands), September (J, H, K-bands) and November (I-band) \citep{dav}. Folding of I--band images on the known orbital period yielded a semi-amplitude of 0.28 $\pm$ 0.17 magnitudes with minimum luminosity at phase 0, consistent with the neutron star being behind the donor. Quiescent observations were also taken using the WHT in Harris-I band (2006 September), which showed short-term flaring of up to 1 magnitude above the average magnitude of 21.83 $\pm$ 0.18 \citep{jon}. This data was folded on the known period and, after removing the flares, a semi-amplitude of $\sim$ 0.06 magnitudes was found, with the system at its brightest at phase 0.34 $\pm$ 0.03. 

Although the 2004 outburst denoted the discovery of 00291+59, it is worth noting that \citet{rem} retrospectively detected `marginal' (5$\sigma$) similar outbursts from the source as detected by the RXTE All-Sky Monitor (ASM) in 1998 November (21 $\pm$  4 mCrab) and 2001 September (18 $\pm$ 3 mCrab) suggesting an outburst from this source every $\sim$ 3 years.

\subsection{The 2008 Outburst}
In 2008 August, 00291+59 again became active. This was first detected by the RXTE PCA on August 13 with a flux of $\sim$ 16 mCrab (2 -- 10 keV). The PCA had been monitoring the source without detection every three days since 2008 May \citep{chak}. A previous observation on August 10 had not detected the source. The August 13 detection was confirmed by Swift X-Ray Telescope (XRT) observations on August 15 \citep{mark08}. Optical photometry was obtained using the Wide Field Camera (WFC) on the Isaac Newton Telescope (INT) \citep{tor08b} and Faulkes Telescope North (FTN) \citep{rus}, noting an increase of \textgreater \ 4 magnitudes from quiescence to values comparable with the 2004 outburst (see Fig. 1). A detection in the UV was made by the Swift UltraViolet/Optical Telescope (UVOT) instrument on August 15 \citep{mars08} in the UVW2 filter (central wavelength, 1928\ \AA). No radio detections were reported, although the WSRT set a 3$\sigma$ upper limit of 0.16 mJy \citep{lin} at 4.9 GHz on August 16. After a rapid fade, a second brightening occurred $\sim$ 1 month later and was detected on September 18, initially in the optical with FTN, subsequently confirmed by Swift X-ray observations \citep{lewb}. This second outburst was just $\sim$ 0.2 magnitudes fainter in i$^\prime$-band than the outburst peak on August 15. 

\begin{figure}[t]
\resizebox{\hsize}{!}{\includegraphics[width=12cm,angle=270]{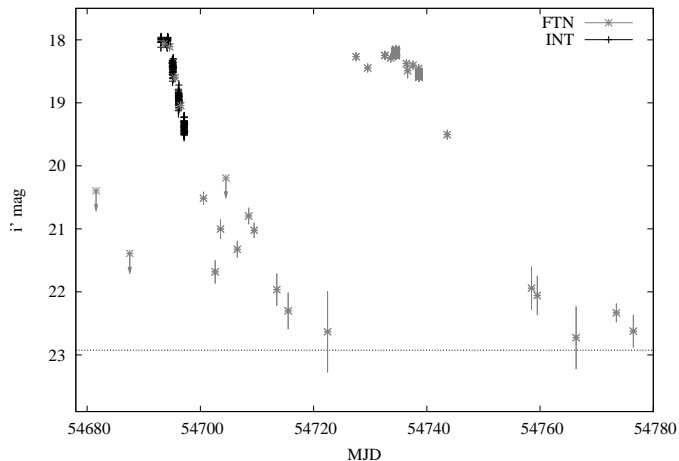}} 
\caption{FT North and INT i$^\prime$-band light curve of the 2008 outburst. The dotted line represents the approximate i$^\prime$-band quiescence level.}
 \label{1}
\end{figure}

Here, we present well-sampled optical light curves of both 2008 outburst peaks, plus a quasi-simultaneous optical spectrum, radio, NIR, UV and X-ray data. We describe the data collection and reduction in Section 2. We analyse the light curve morphology, the source's optical variability, spectral properties and broadband behaviour in Sections 3 -- 5. We present the evolution of the NIR--optical--UV SEDs (which are the most complete of any AMXP to date) and the first broadband (radio--X--ray) SEDs of 00291+59, and compare this outburst to the 2004 outburst and to other AMXPs in Section 6. Our conclusions are presented in Section 7.

\section{Observations and Data Reduction} 

 \subsection{Optical, UV and Near Infrared Data}
Data are presented from Faulkes Telescope North (optical), Swift UVOT (UV/optical), INT (optical), an optical spectrum from Keck and PAIRITEL (NIR). The photometric observations are listed in Table 1.

To convert the magnitudes to intrinsic de-reddened flux densities, we use A$_{V}$ = 2.5 $\pm$ 0.3 for the interstellar extinction towards the source \citep{tor08a} and use the extinction curve of \citet{car}. The resulting flux densities will have absolute uncertainties due to the errors in the value of the extinction, corresponding (since extinction increases at shorter wavelengths) to 3\%, 28\% and 109\% of the flux at K, V and UVW2 wavelengths respectively (adopting values of A$_{K}$ = 0.114 and A$_{UVW2}$ = 2.664 from Table 3 and Equation 4 in \citealp{car}). These errors are taken into account for the SEDs presented in Section 6. 

\begin{table}
\caption{Photometric observations (NIR, optical and UV detections) of 00291+59 used in this work.}
\begin{tabular}{lccl}
\hline
Telescope &UT Date   &MJD    &Exposure times\\
/ detector&          &       &per filter    \\
\hline
INT/WFC      &2008-08-15&54693.0&120 r$^\prime$,i$^\prime$ (10), 30 i$^\prime$ (2)\\
FTN/EA01,EM01&2008-08-15&54693.6&200 B,V,R,i$^\prime$,y \\
Swift/UVOT   &2008-08-15&54693.7&1462 UVW2$^{\ast}$ \\
INT/WFC      &2008-08-16&54694.2&120 r$^\prime$,i$^\prime$ (30), 30 r$^\prime$ \\
FTN/EA01,EM01&2008-08-16&54694.6&200 B,V,R,i$^\prime$,y \\
INT/WFC      &2008-08-17&54695.1&120 r$^\prime$ (30), 120i$^\prime$ (48),\\
&&&30 r$^\prime$ \\
FTN/EA01,EM01&2008-08-17&54695.5&200 B,V,R,i$^\prime$,y \\
INT/WFC      &2008-08-18&54696.2&120 i$^\prime$ (50) \\
FTN/EA01,EM01&2008-08-18&54696.5&200 B,V,R,i$^\prime$,y \\
INT/WFC      &2008-08-19&54697.1&120 i$^\prime$ (48) \\
FTN/EA01     &2008-08-22&54700.5&200 i$^\prime$ (19)$^{\ast}$ \\
FTN/EM01     &2008-08-24&54702.6&200 B,V,R,i$^\prime$ \\ 
FTN/EM01     &2008-08-25&54703.6&200 R,i$^\prime$ \\
FTN/EM01     &2008-08-28&54706.5&200 B,V,R,i$^\prime$,y \\
Swift/UVOT   &2008-08-29&54707.9&1484 B$^{\ast}$ \\
FTN/EM01     &2008-08-30&54708.5&200 i$^\prime$ \\
FTN/EM01     &2008-08-31&54709.5&200 i$^\prime$ \\
FTN/EM01     &2008-09-04&54713.5&200 i$^\prime$ \\
FTN/EM01     &2008-09-06&54715.5&200 i$^\prime$ \\
FTN/EM01     &2008-09-13&54722.5&200 i$^\prime$ \\
FTN/EM01     &2008-09-18&54727.5&200 i$^\prime$ \\
Swift/UVOT   &2008-09-20&54729.1&115 B$^{\ast}$, 715 UVW2$^{\ast}$\\
FTN/EM01     &2008-09-20&54729.5&200 i$^\prime$ \\
PAIRITEL/2M$^{\dagger}$&2008-09-21&54730.3&188 J$^{\ast}$, 730 H$^{\ast}$,K$^{\ast}$ \\
PAIRITEL/2M$^{\dagger}$&2008-09-23&54732.3&1107 J$^{\ast}$,H$^{\ast}$,K$^{\ast}$ \\
FTN/EM01     &2008-09-23&54732.6&200 V,i$^\prime$ (2), 200 R \\
FTN/EA01,EM01&2008-09-24&54733.6&200 V,R,i$^\prime$ \\
PAIRITEL/2M$^{\dagger}$&2008-09-25&54734.3&259 J$^{\ast}$,H$^{\ast}$,K$^{\ast}$ \\
FTN/EA01,EM01&2008-09-25&54734.5&200 i$^\prime$ (44) \\
FTN/EA01,EM01&2008-09-27&54736.5&200 R,i$^\prime$ (2),B,V \\
FTN/EA01,EM01&2008-09-28&54737.5&200 B,V,R,i$^\prime$ \\
FTN/EM01     &2008-09-29&54738.6&200 i$^\prime$ (22) \\
PAIRITEL/2M$^{\dagger}$&2008-10-01&54740.3&1601 J$^{\ast}$,H$^{\ast}$,K$^{\ast}$ \\
FTN/EM01     &2008-10-04&54743.6&200 B,V,R,i$^\prime$ \\
PAIRITEL/2M$^{\dagger}$&2008-10-06&54745.2&47 J$^{\ast}$,1224 H$^{\ast}$ \\
FTN/EM01     &2008-10-19&54758.5&200 i$^\prime$ \\
FTN/EM01     &2008-10-20&54759.5&200 R,i$^\prime$ \\
FTN/EM01     &2008-10-27&54766.4&200 i$^\prime$ \\
FTN/EM01     &2008-11-03&54773.5&200 i$^\prime$ (10)$^{\ast}$ \\
FTN/EM01     &2008-11-06&54776.5&200 i$^\prime$ \\
\hline
\end{tabular}
\small
\\ MJD = Modified Julian Day. $^{\ast}$combined (aligned and stacked) to produce one image; $^{\dagger}$2MASS Survey cam. Numbers in parentheses denote multiple exposures, e.g. 120 r$^\prime$,i$^\prime$ (10) is 10 exposures of 120 seconds in each of r$^\prime$ and i$^\prime$. For details of telescopes, cameras and filters, see respective sections of text, 2.1.1, 2.1.2, 2.1.3 and 2.1.5.
\normalsize
\end{table}
 \subsubsection{Faulkes Telescope North}  

00291+59 has been included as part of a monitoring campaign of 30 LMXBs including 5 AMXPs \citep{lewa} using the two Faulkes Telescopes. Data have been collected using the 2-metre robotic Faulkes Telescope North located at Haleakala on Maui. It currently uses a  Merope camera (EM01). Prior to 2008 August, the camera used was the Apogee `Hawkcam' (EA01). Both cameras were coupled with an E2V CCD42-40DD CCD giving a 4\farcm7 $\times$ 4\farcm7 field of view and producing images of 2048 $\times$ 2048 pixels binned 2 $\times$ 2 to give 1024 $\times$ 1024 pixels at 0.278 arcsecond pixel $^{-1}$. Science images are produced using the Faulkes automatic pipeline which de-biases and flat-fields the raw images. Filters used are B, V, R (Bessell), i$^\prime$ (Sloan-SDSS) and y (Pan-STARRS). We monitored the source in i$^\prime$-band, typically acquiring a 200 second exposure every $\sim$ 2 weeks when the source was visible from FTN. When the outburst was detected on 2008 August 1, we increased the sampling to one observation every 2 days and used B,V, R, i$^\prime$, y  filters to investigate colour changes. This shorter cadence continued until November 10, near the end of the fade of the second outburst peak.

Seeing values range from 0\farcs4 \ to $\sim$ 3\farcs1. Images were discarded if the signal-to-noise ratio was low (often due to thin cloud), or if the tracking or focus were poor. During the outburst, we detected the source in 133 images taken between 2008 August and November. We performed aperture photometry of 00291+59 and two nearby field stars (see Table 2 and Fig. 2) using the aperture photometry package \small APPHOT \normalsize in \small IRAF \normalsize. Point-spread-function (PSF) fitting was not used since minor errors in tracking on some observations resulted in non-circular or non-elliptical stars for which aperture photometry was more suitable. A fixed aperture radius of 6 pixels along with a background annulus of 10 -- 20 pixels was adopted for all three stars in all filters.

Flux calibration in B, V and R-bands was achieved using photometry of the standard star field PG 0231+051 from the list of Landolt photometric stars \citep{lan}, which are observed regularly by FTN. Both the standard and the 00291+59 fields were observed on 2008 September 4, 6 and 14. Four stars in the same field as PG 0231+051 with known B, V, R magnitudes \footnotemark \footnotetext[1]{http://eso.org/sci/facilities/paranal/sciops/Bessell/PG0231+051.ps} were also used. Accounting for differences in airmass between target field and standard, we calculated the B, V, R magnitudes of the two chosen field stars in the 00291+59 field. From the three dates used, the measured magnitudes varied by only $\sim 0.01$ mag, indicating the conditions were photometric on all three dates. Magnitude errors were estimated from the (night-to-night) range of measured magnitudes of the two field stars and measured differences in the relative magnitudes of the stars in the PG 0231+051 field.

For the i$^\prime$-band flux calibration, the field centred on RA 23:48:20, dec +00:57:18 (J2000.0) was observed 9 times on 2008 September 27, at airmasses from 1.58 -- 2.04. Instrumental magnitudes were measured for the star SDSS J234817.22+005557.2 and its values compared with that given in the Sloan Digital Sky Survey (SDSS) Data Release 6 (DR6). Comparisons were made between this star and others in the field to ensure that it was non-variable and that conditions remained photometric. Having calculated the change in offset (between instrumental and known magnitudes) with airmass, and accounting for the difference in exposure times between this and the target field, we were able to derive magnitudes for stars 1 and 2 as shown in Table 2. The uncertainties include the instrumental uncertainties given by \small IRAF \normalsize for stars 1, 2 and SDSS J234817.22+005557.2, summed in quadrature with the value quoted for SDSS J234817.22+005557.2 in SDSS DR6.

Spectroscopic standards were required for the y-band calibration, since magnitudes in this Pan-STARRS y filter (which has an effective wavelength of 1.004 $\mu$m) are not typically known for photometric standard stars. The spectroscopic standards BD+28 4211 and Feige 110 \citep{oke} were both used on October 27 (4 and 5 exposures, respectively) and October 31 (one exposure of each). Both standards have smooth spectra with known AB magnitudes in steps of 2\ \AA. Using the same method as for the i$^\prime$-band, we estimated the y-band magnitudes of the two 00291+59 field stars. The uncertainties incorporate the small differences in measurements between the two dates. 

\begin{figure}[t]
\resizebox{\hsize}{!}{\includegraphics[width=12cm,angle=0]{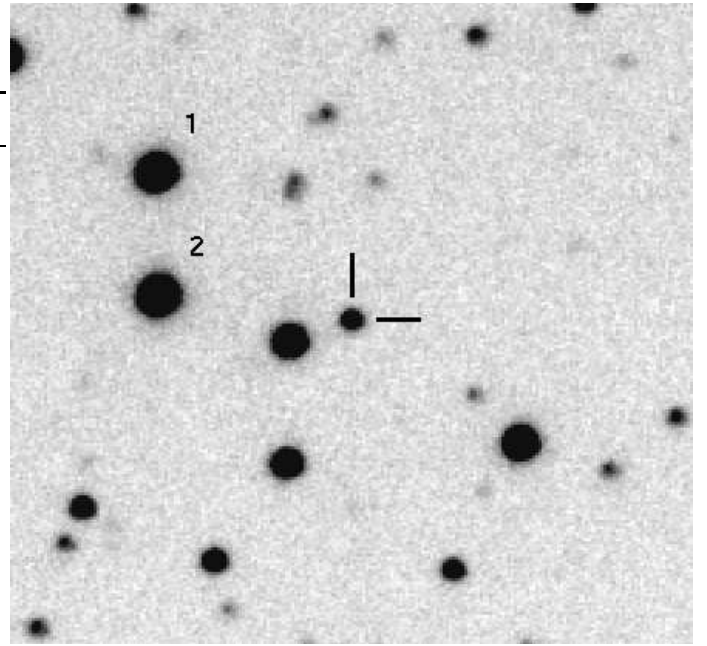}} 
\caption{FT North finder chart (SDSS-i$^\prime$) for IGR J00291+5934 from the 2008 outburst. North is up and East is to the left. Image was taken on 2008 September 20, exposure 200 seconds, image size is 1\arcmin $\times$ 1\arcmin. The magnitude of the target in the image is i$^\prime$ = 18.45 $\pm$ 0.07.}
 \label{2}
\end{figure}

\begin{table}
\caption{Optical and NIR magnitudes of the two comparison stars in the 00291+59 field used for flux calibration.}
\begin{tabular}{ccccl}
\hline
Star&1&2&Reference \\
RA&00h29m05.38s&00h29m05.3s&\\
Dec&+59d34m32.17s&+59d34m20.9s&\\
\hline
B &17.03 $\pm$ 0.04&16.51 $\pm$ 0.04&FTN; this paper \\
V &16.02 $\pm$ 0.02&15.60 $\pm$ 0.02&FTN; this paper \\
r$^\prime$&15.41 $\pm$ 0.01&15.07 $\pm$ 0.01&\citealp{tor08a} \\
R &15.39 $\pm$ 0.05&15.07 $\pm$ 0.05&FTN; this paper \\
i$^\prime$&15.46 $\pm$ 0.02&15.22 $\pm$ 0.02&FTN; this paper \\
y &14.81 $\pm$ 0.14&14.72 $\pm$ 0.13&FTN; this paper \\
J &13.80 $\pm$ 0.02&13.75 $\pm$ 0.02&2MASS\\
H &13.39 $\pm$ 0.03&13.29 $\pm$ 0.03&2MASS\\
K &13.24 $\pm$ 0.03&13.24 $\pm$ 0.03&2MASS\\
\hline
\end{tabular}
\small
\\2MASS is the Two Micron All Sky Survey.
\normalsize
\end{table}

  \subsubsection{Swift UVOT}

The Swift UVOT observed 00291+59 on nine dates in 2008. Four detections (at 5$\sigma$) of the source were made during the 2008 outburst (see Tables 1 and 3); two in B-band and two in UVW2, which has an effective wavelength of 1928\ \AA. Individual images were combined and magnitudes and upper limits were derived using the standard Swift UVOT routines provided by NASA's High Energy Astrophysics Science Archive Research Center (HEASARC). Apertures of 3\arcsec were used centred on the known coordinates of 00291+59. As a confirmation, we measured the B magnitude of field star 1 to be consistent with that measured by FTN (see Table 2) to an accuracy of $< 0.1$ mag. On some dates, $> 10$ images were acquired in the same filter, but the counts of 00291+59 were never high enough to produce a light curve. In Table 3 we list the UVOT magnitudes when detected and upper limits for non-detections.

\begin{table}
\caption{Swift UVOT optical/UV magnitudes and 3$\sigma$ upper limits of 00291+59.}
\begin{tabular}{ccl}
\hline
Date&MJD&magnitudes\\
\hline
2008-08-15&54693.7&UVW2 = 19.70 $\pm$ 0.11 \\
2008-08-21&54699.9&B $>$ 20.57; UVW2 $>$ 20.40 \\
2008-08-23&54701.9&UVW2 $>$ 20.13 \\
2008-08-27&54705.6&V $>$ 18.63; B $>$ 20.71; U $>$ 20.31 \\
          &       &UVW1 $>$ 20.49; UVM2 $>$ 20.36\\
          &       &UVW2 $>$ 20.98 \\
2008-08-29&54707.9&B = 21.13 $\pm$ 0.19 \\
2008-08-30&54708.6&UVW2 $>$ 21.48 \\
2008-09-04&54713.5&UVW1 $>$ 21.44 \\
2008-09-20&54729.1&B = 19.22 $\pm$ 0.16; UVW1 $>$ 18.62 \\
          &       &UVW2 = 20.11 $\pm$ 0.21 \\
2008-10-23&54762.9&UVM2 $>$ 21.41 \\
\hline
\end{tabular}
\normalsize
\end{table}

  \subsubsection{INT}
Observations of 00291+59 were taken using the WFC at the 2.5 metre Isaac Newton Telescope (INT) at Roque de los Muchachos, La Palma, Spain, between 2008 August 15 and 19. The WFC has an image scale of 0.333 arcsecond pixel $^{-1}$. Observations were made using the Sloan r$^\prime$ and i$^\prime$ filters. A total of 188 observations were taken in i$^\prime$-band over the 5 nights, with a further 92 taken in r$^\prime$-band over the first 3 nights. A  master bias and flat-field were created and used to reduce the raw images. Photometry was performed in the same way as for the FTN data (Section 2.1.1) with the same comparison stars, apertures and annuli (in pixels). Flux calibration for i$^\prime$-band was performed as for the FTN i$^\prime$-band data; for r$^\prime$-band, the calibration values of \citet{tor08a} were used (see Table 2). 
 
\begin{figure*}[t]
 \centering
 \resizebox{\hsize}{!}{\includegraphics[width=12cm,angle=270]{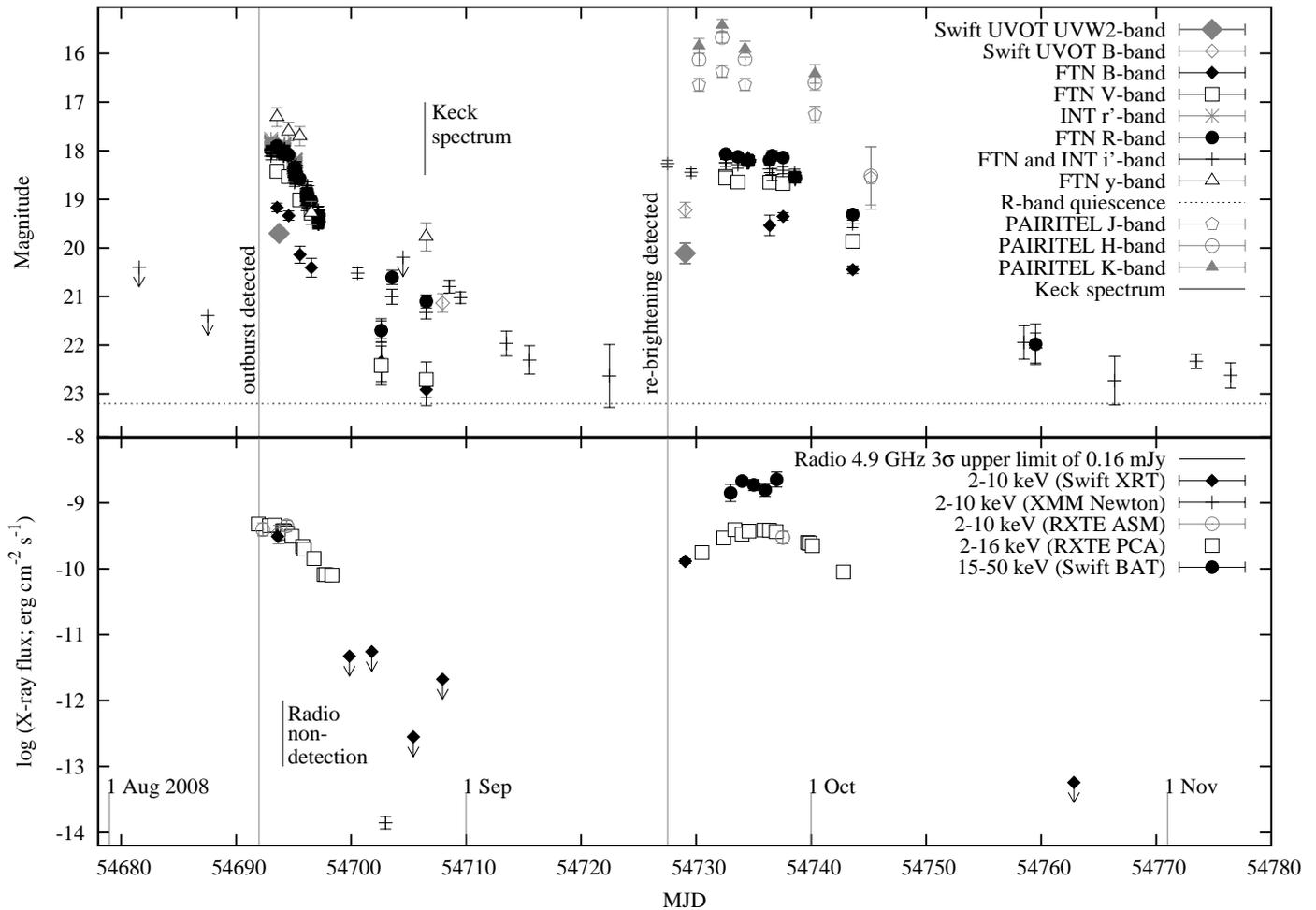}} 
 \caption{\emph{Upper}: Optical, UV and NIR data from the 2008 outburst. \emph{Lower}: X-ray and radio data from the same period. We also plot earlier FTN i$^\prime$-band non-detections and the date of the Keck spectrum in the upper panel and the radio and Swift XRT non-detections in the lower panel.}
 \label{3}
\end{figure*}

\begin{table}
\caption{PAIRITEL NIR magnitudes and upper limits of 00291+59.}
\begin{tabular}{ccccc}
\hline
Date&MJD&J&H&K\\
\hline
2008-09-21&54730.3&16.65 $\pm$ 0.13&16.13 $\pm$ 0.13&15.85 $\pm$ 0.15 \\
2008-09-23&54732.3&16.37 $\pm$ 0.12&15.67 $\pm$ 0.11&15.42 $\pm$ 0.13 \\
2008-09-25&54734.3&16.64 $\pm$ 0.12&16.12 $\pm$ 0.12&15.91 $\pm$ 0.17 \\
2008-10-01&54740.3&17.26 $\pm$ 0.17&16.60 $\pm$ 0.15&16.42 $\pm$ 0.19 \\
2008-10-06&54745.2&18.56 $\pm$ 0.64&18.52 $\pm$ 0.60&$>$ 17.09 \\
2008-10-07&54746.2&$>$ 17.30&$>$ 16.90&$>$ 17.60 \\
2008-10-08&54747.2&$>$ 17.39&$>$ 16.79&$>$ 17.49 \\
2008-10-13&54752.2&$>$ 17.33&$>$ 16.79&$>$ 17.40 \\
\hline
\end{tabular}
\normalsize
\end{table}

 \subsubsection{Keck Spectrum} 

A single spectrum was acquired on 2008 Aug  28 using the Low Resolution Imaging Spectrometer (LRIS) attached to the Keck 1 telescope. 00291+59 was observed using a  1\farcs0 wide slit with the CCDs on each spectrograph arm unbinned. The exposure time for both blue and red arms was 900 seconds. The red arm was used with the 400/8500 grating. The useful wavelength range covered the 5470 -- 9240\ \AA\ interval with a dispersion of 1.86\ \AA\ per pixel and a spectral resolution of about 6\ \AA\ FWHM.  The blue arm was used with the 600/4000 grism to yield an useful wavelength coverage of 3890 -- 5600\ \AA\ with a dispersion of 0.62\ \AA\ per pixel. The blue arm data were too noisy to obtain a meaningful spectrum. Conditions were not considered to be photometric; seeing was variable from 0\farcs7 -- 1\farcs0. The spectrum was extracted and reduced using \small KPNOSLIT \normalsize in \small IRAF \normalsize with 1\arcsec slit and using the standard star, Feige 110. We use the Starlink page \small DIPSO \normalsize to measure equivalent widths.

 \subsubsection{PAIRITEL}

NIR photometry was obtained with the 1.3m Peters Automated Infrared Imaging Telescope (PAIRITEL) at Fred Lawrence Whipple Observatory (FLWO) \citep{blo} which images simultaneously a 8\farcm5 $\times$ 8\farcm5 field of view in the J, H and Ks photometric bands.  The observations consisted of a large number of dithered 7.8 s exposures on source. The dithered exposures were first bias and flat-field corrected to be mosaiced together for each individual visit (see e.g. \citealp{bla}). The observations were made on 8 nights between 2008 September 21 and October 13, although observations after October 6 yielded non-detections. Photometry was performed using the same comparison stars and method as previously described, however with apertures of radius 3 pixels to compensate for the greater image scale. The J, H, K 2MASS magnitudes of the two field stars (Table 2) were used to calibrate the flux of 00291+59. The magnitudes and upper limits are given in Table 4.

 \subsection{X-ray Data}
We use pointed observations from the Swift XRT, XMM-Newton and RXTE PCA and augment these with publicly available data from Swift Burst Alert Telescope (BAT) and RXTE ASM. We obtain a light curve from the RXTE PCA data (14 detections in August and 13 in September). We list the X-ray observations in Table 5.

\begin{table*}
\caption{X-ray observations of 00291+59 used in this work.}
\begin{tabular}{lccccl}
\hline
Telescope / Detector&UT Date      &MJD    &Energy range (keV)&Exposure times (ksec)&Observation type\\
\hline
RXTE/PCA      &2008-08-13&54691.9&2--16 &1.0  &Pointed\\
RXTE/ASM      &2008-08-14&54692.3&1.5--12 &--   &Monitoring 3$\sigma$ detection\\
RXTE/PCA      &2008-08-14&54692.9&2--16 &3.8  &Pointed\\
RXTE/PCA      &2008-08-15&54693.3&2--16 &0.9  &Pointed\\
Swift/XRT     &2008-08-15&54693.6&0.5--10  &1.9  &Pointed (PC mode)\\
RXTE/ASM      &2008-08-15&54693.7&1.5--12 &--   &Monitoring 3$\sigma$ detection\\
RXTE/ASM      &2008-08-16&54694.4&1.5--12 &--   &Monitoring 3$\sigma$ detection\\
RXTE/PCA      &2008-08-16&54694.5&2--16 &28.1 &5 $\times$ Pointed\\
RXTE/PCA      &2008-08-17&54695.8&2--16 &5.9  &2 $\times$ Pointed\\
RXTE/PCA      &2008-08-18&54696.8&2--16 &2.7  &Pointed\\
RXTE/PCA      &2008-08-19&54697.7&2--16 &7.4  &2 $\times$ Pointed\\
RXTE/PCA      &2008-08-20&54698.3&2--16 &2.2  &Pointed\\
Swift/XRT     &2008-08-21&54699.9&0.5--10   &1.2  &Pointed (WT mode)\\
Swift/XRT     &2008-08-23&54701.8&0.5--10   &0.4  &Pointed (WT mode)\\
XMM/MOS       &2008-08-25&54703.4&0.5--10   &33   &Pointed\\
Swift/XRT     &2008-08-27&54705.4&0.5--10   &2.2  &Pointed (PC mode)\\
Swift/XRT     &2008-08-29&54707.9&0.5--10   &4.9  &Pointed (WT mode)\\
Swift/XRT     &2008-09-20&54729.0&0.5--10   &1.0  &Pointed (PC mode)\\
RXTE/PCA      &2008-09-21&54730.5&2--16 &0.9  &Pointed\\
RXTE/PCA      &2008-09-23&54732.4&2--16 &1.2  &Pointed\\
Swift/BAT     &2008-09-24&54733.0&15--50  &--   &Monitoring 3$\sigma$ detection\\
RXTE/PCA      &2008-09-24&54733.7&2--16 &2.0  &2 $\times$ Pointed\\
Swift/BAT     &2008-09-25&54734.0&15--50  &--   &Monitoring 3$\sigma$ detection\\
RXTE/PCA      &2008-09-25&54734.6&2--16 &11.2 &Pointed\\
Swift/BAT     &2008-09-26&54735.0&15--50  &--   &Monitoring 3$\sigma$ detection\\
RXTE/PCA      &2008-09-26&54735.9&2--16 &2.5  &Pointed\\
Swift/BAT     &2008-09-27&54736.0&15--50  &--   &Monitoring 3$\sigma$ detection\\
RXTE/PCA      &2008-09-27&54736.7&2--16 &4.9  &2 $\times$ Pointed\\
Swift/BAT     &2008-09-28&54737.0&15--50  &--   &Monitoring 3$\sigma$ detection\\
RXTE/ASM      &2008-09-28&54737.5&1.5--12 &--   &Monitoring 3$\sigma$ detection\\
RXTE/PCA      &2008-09-30&54739.8&2--16 &4.6  &3 $\times$ Pointed\\
RXTE/PCA      &2008-10-01&54740.1&2--16 &0.7  &Pointed\\
RXTE/PCA      &2008-10-03&54742.8&2--16 &10.2 &Pointed\\
Swift/XRT     &2008-10-23&54762.8&0.5--10   &2.0  &Pointed (PC mode)\\
\hline
\end{tabular}
\small
\normalsize
\end{table*}

\subsubsection{Swift XRT}
We analysed all Swift XRT observations of the field taken between 2008 August and October. We performed the standard screening on all observations using the latest `xrtpipeline' (v. 0.12.3). In each of the two observations where 00291+59 was detected we extracted a source spectrum. In both cases, the XRT was in photon counting (PC) mode and the observations suffered from pile up. To correct for this, following the Swift Science Data Centre (SSDC) recommendations 
\footnotemark \footnotetext[2]{http://www.swift.ac.uk/pileup.shtml}, we used an annular extraction region centred on 00291+59's position with inner radii of $\sim$ 20\arcsec~and $\sim$ 10\arcsec~for observations 00031253001 and 00031253005, respectively.

We thereby excluded the core of the PSF from our analysis, resulting in encircled energy fractions at the inner radius of $\sim$ 50\% and $\sim$ 40\% for 00031253001 and 00031253005, respectively. The annuli's outer radius was fixed at $\sim$ 120\arcsec. Background spectra were accumulated far from the source using regions of the same size. We created exposure maps and ancilliary files using `xrtexpomap' (v.0.2.5) and `xrtmkarf' (v.0.5.6), respectively, thereby accounting for the energy lost in the PSF core. We grouped the spectra to have a minimum of 20 counts per energy bin and fitted them within Xspec (v. 11.3.2ag) using the latest available response matrices (swxpc0to12s6\_20010101v011.rmf) and the 0.5 -- 10 keV energy range. The spectra were satisfactorily fitted (reduced $\chi$$^{2}$ of 0.9 for 77 and 33 degrees of freedom) with a simple absorbed power law model. The measured fluxes, absorption and power law parameters are given in Table 6, along with 3$\sigma$ flux upper limits when the source was not detected.

 \subsubsection{XMM-Newton}

We analysed the X-ray Multi-Mirror satellite (XMM-Newton) observation of 00291+59 taken on MJD 54703 (2008 August 25), which lasted approximately 33 kiloseconds. 00291+59 was clearly detected in both the MOS1 and MOS2 detectors, which operated in large window (medium filter) mode, at an average net rate of $\sim$ 0.01 counts second$^{-1}$. We excluded contaminating flares from the observation, defined as segments with a count rate above 10 keV higher than 1 count second$^{-1}$. We then defined source and background circular regions, both with a radius of 30\arcsec. We took into account the extraction area using the tool `backscale' and created response matrices and ancilliary files with `rmfgen' and `arfgen', respectively. We grouped the channels in the resulting spectra in order to have a minimum of 15 counts per energy bin, and fitted them within Xspec using the 0.5 -- 10 keV energy range, using an absorbed power law model. We note that, as with the majority of the Swift XRT observations (see Section 2.1.1), the low count rates from the source preclude us from studying variability in detail within the observation window, although we note that the level of variability is consistent with the source being at a constant flux.

 \subsubsection{RXTE PCA}

The PCA was used to monitor the X-ray flux during both outburst peaks \citep{chak,gall08}. We use the 16 sec time resolution Standard 2 mode data to calculate the X-ray flux in the 2 -- 16 keV energy band. The energy-channel conversion is done by using the pca\_e2c\_e05v02 table provided by the RXTE team. The deadtime was corrected and the number of background events within our energy range and time interval was estimated using the FTOOL `pcabackest', following the standard procedure suggested by the RXTE manual \footnotemark \footnotetext[3]{http://heasarc.gsfc.nasa.gov/docs/xte/recipes/pcabackest.html}. We reject all data with a measured flux (2 -- 16 keV) of $F < 8 \times 10^{-11}$ erg cm$^{-2}$ s$^{-1}$ because flux from background contamination was likely significant. This is most evident from an apparent PCA detection on MJD 54700 of $F \sim 6.1 \times 10^{-11}$ erg cm$^{-2}$ s$^{-1}$ whereas a Swift XRT pointing fours hours earlier did not detect the source, with an upper limit of $4.7 \times 10^{-12}$ erg cm$^{-2}$ s$^{-1}$. The nearby intermediate polar V709 Cas is in the PCA's field of view, and contributes some contaminating flux (see \citealp{lin07}) however the flux level of this source is lower than the above flux limit, and therefore its contribution is minimal.

 \subsubsection{Publicly Available Data From X-ray Monitors}
Five 3$\sigma$ detections were also made by the Swift BAT instrument (15 -- 50 keV). All five detections were during the second peak, in 2008 September; BAT did not detect the source during the first outburst peak. We include these public data in our analysis. We also include four 3$\sigma$ detections from the public archive of the RXTE ASM (1.5 -- 12 keV); three from the first outburst peak and one from the second. All RXTE ASM fluxes were converted to unabsorbed (adopting $n_{\rm H} = 4.64 \times 10^{21}$ cm$^{-2}$ from \citealp{tor08a}) 2 -- 10 keV fluxes using the HEASARC tool WebPIMMS to compare with the Swift XRT and XMM 2 -- 10 keV data. A power law index of 1.6 (as measured by Swift; see Table 6) was adopted.

 \subsection{WSRT Radio Data}
00291+59 was observed with the Westerbork Synthesis Radio Telescope on 2008 August 15 -- 16 \citep{lin}. The observation was made between $\sim$ 20:30 -- 06:30 UT at the median frequency of 4.9 GHz, with a total bandwidth of 160 MHz. The primary calibrator used was 3C 286. The calibration and analysis of the data were done using \textsc{Miriad} \citep{sau}.
 
No object was detected at the position of the target reported by \citet{rup}. The 3$\sigma$ upper limit to the flux density (measured in the image plane) was 0.16 mJy. 

\section{Analysis of the 2008 Outburst}
The multi-wavelength light curve of the 2008 outburst is shown in Fig. 3. The source displays two separate periods of activity, which we describe as `peaks' within the overall `outburst' (Fig.1). We display the initial X-ray detections (MJD 54691) of the outburst and its subsequent detection in the optical/infrared (OIR) at MJD 54693. FTN observations from long-term monitoring \citep{lewa} had provided non-detections, most recently on August 3 and 9 (i$^\prime$ \textgreater 20.40 and 21.39 respectively), showing that the OIR brightened by at least 0.5 mag day$^{-1}$. We show a well-sampled UV--IR fade over the following 30 days accompanied by RXTE PCA and Swift XRT monitoring. For the first three days, the X-ray had an approximately constant flux (the optical flux also seemed approximately constant) before the source faded, more rapidly in X-ray than in OIR. We present data from Swift XRT observations in Table 6. The XRT X-ray spectra are hard, and can be fit with an absorbed power law with a photon index typical of AMXPs in outburst. Just 11 days after the first X-ray detection with Swift XRT, an XMM-Newton detection showed it had drastically faded. From XMM simultaneous fitting of MOS 1 and 2 spectra using the `wabs*powerlaw' model, we derive an unabsorbed 2 -- 10 keV flux of 1.4 $\pm$ 0.3 $\times$ 10$^{-14}$ erg cm$^{-2}$ s$^{-1}$, i.e. more than four orders of magnitude lower than the outburst peak fluxes. Further details of the XMM spectrum will be reported in Linares et al. (in preparation). We note that long after the source was no longer detectable with Swift, the source remained visible in the OIR (i$^\prime$-band), eventually approaching its quiescent value at $\sim$ MJD 54722 (31 days after the outburst was first detected).

\begin{figure}[t]
 \includegraphics[width=6.4cm,angle=270]{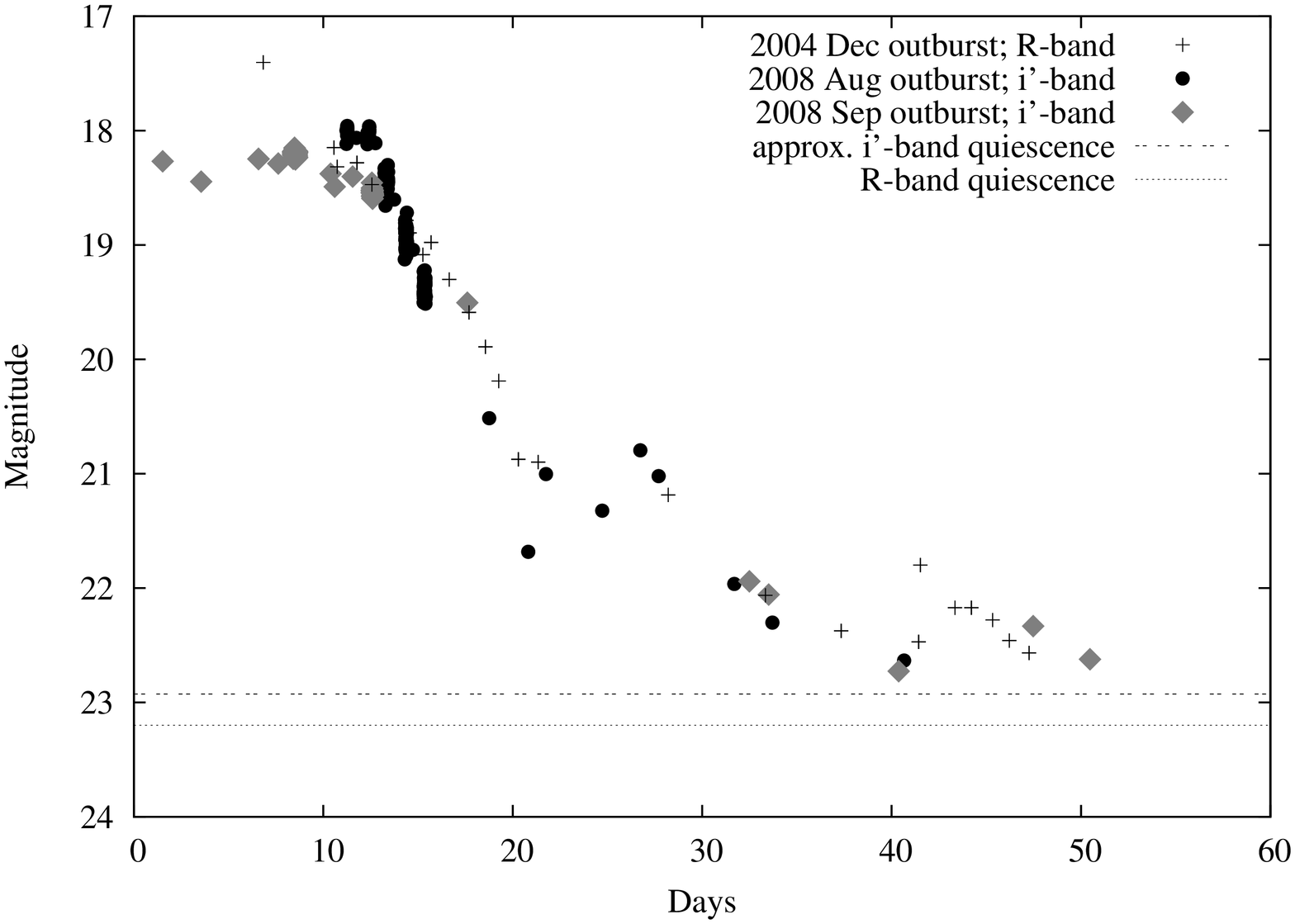}
 \includegraphics[width=6.4cm,angle=270]{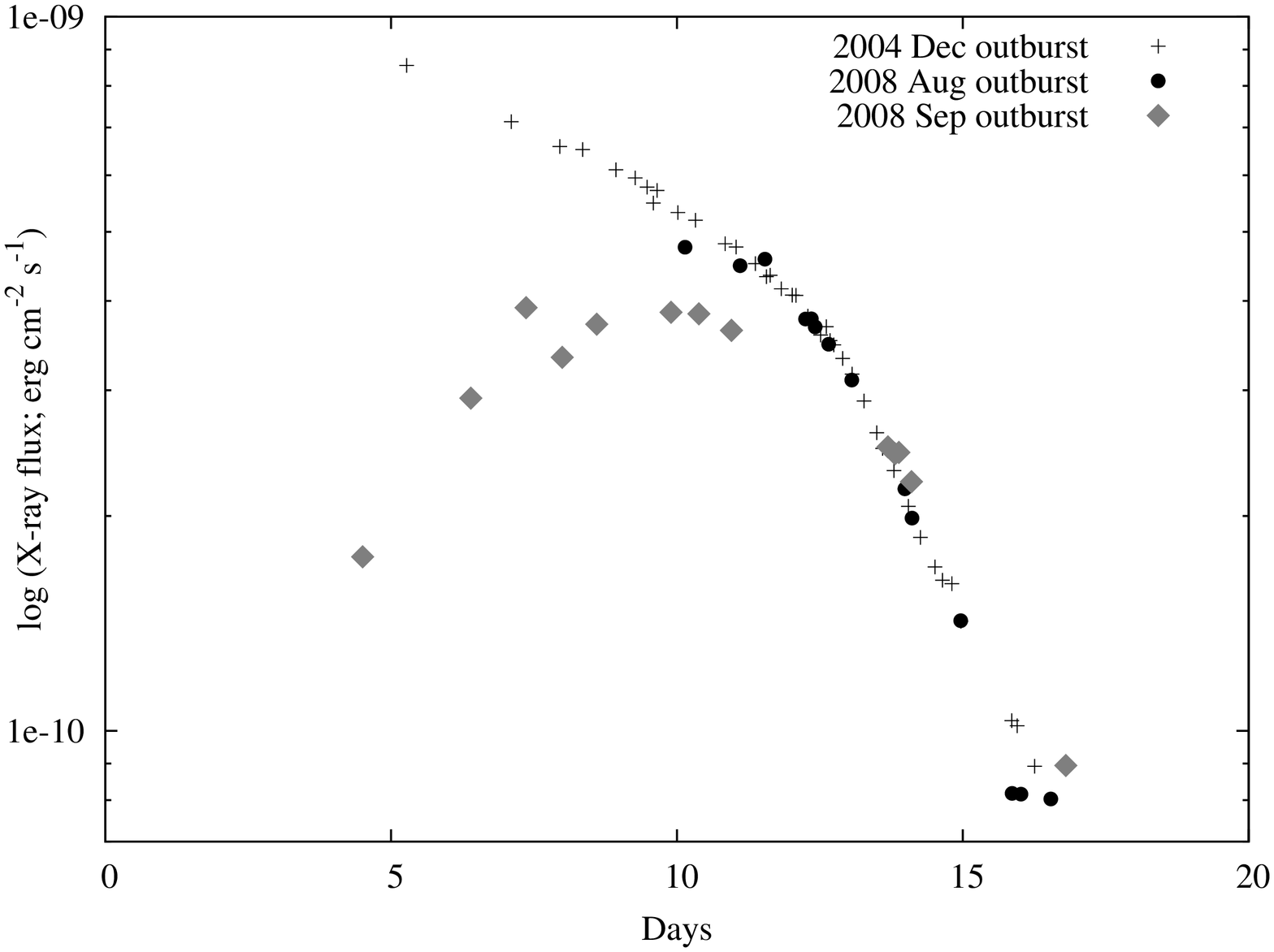}\\
 \caption{\emph{Upper}: Comparison of the three fades using R-band data from 2004 \citep{tor08a} and i$^\prime$-band data from FT North. \emph{Lower}: The same for the 2 -- 16 keV X-ray flux from the RXTE PCA (note the different time ranges).}
 \label{4}
\end{figure}

A second `peak' was detected; this time, first in optical \citep{lewb}, with the source re-brightening by 4 -- 5 magnitudes to $\sim$ 18.3 (MJD 54727) in i$^\prime$-band, very similar to the magnitude of the first peak in August. It is clear that the 2008 August and September peaks differ, with the August peak fading rapidly and the September peak remaining in a bright `plateau' at near maximum luminosity for $\sim$ 10 days. This second peak lasted $\sim$ 70\% longer than its predecessor (i.e.$\sim$ 50 days as opposed to $\sim$ 30 days). Whereas the i$^\prime$-band light curve appeared fairly constant during the September plateau, the X-ray and NIR light curves brightened then faded slightly, before eventually all bands faded towards quiescence. In the NIR, the 2004 outburst was estimated to reach a maximum of K $\lesssim$ 16 \citep{ste} from a quiescent K $\sim$ 19; in 2008, we measure (during the second peak) K $\sim$ 15.5. This is slightly at odds with the evidence in the R-band which suggests the 2008 outburst was slightly fainter than that of 2004, however we note that the 2004 K-band magnitude was from observations 4 days after the initial detection.

From Fig. 3 and Table 5, we can see that the RXTE ASM (2 -- 10 keV) detected the first peak on three occasions, and the second peak only once. In contrast, Swift BAT  (15 -- 50 keV) did not detect the first peak, and yet detected the second peak on four occasions. However, we note that the respective sensitivities of the ASM and BAT are variable in time, and indeed we see that the photon indices as measured by Swift XRT were consistent with being the same at the start of both peaks (see Table 6). In Fig. 4, we plot data for the 2004 outburst alongside the two separate peaks of the 2008 outburst, for the optical (upper panel) and X-ray (lower panel) fades. The 2004 outburst was measured at its maximum at R  $\sim$ 17.4 \citep{fox}, as compared with the 2008 outburst, which reached values of R $\sim$ 17.9. In the optical, there is a rapid fade between days $\sim$ 12 -- 20 after all three peaks and the same is evident in X-ray up to at least day 16. It is remarkable how identical the three X-ray light curve decays are, given that the initial phases (before day 10 in Fig. 4) are so different. The light curves are overlaid such that these fades are aligned (the same shifts are applied in X-ray and optical). Within the fade from the first peak, we see possible evidence of optical reflares of $\sim$ 1 magnitude ($\sim$ day 20 -- 30, Fig.4 ). This post-peak flaring activity has only been observed from one other AMXP; SAX J1808.4$-$3658 in its X-ray light curve, where the system flares by several orders of magnitude weeks after the initial outburst \citep{patruno}. 

The 2008 August fade from peak to quiescence (in i$^\prime$-band) is measured as $\sim$ 4.6 magnitudes over a period of 29 days giving an average fade of 0.16 $\pm$ 0.02 magnitudes day$^{-1}$. The initial fade (from day 2 to day 5 of the outburst) is steeper with a fitted value of 0.45 $\pm$ 0.01 magnitudes day$^{-1}$. The rise to the second peak over $\leq $ 5 days was measured as $\geq$ 0.72 magnitudes day$^{-1}$. This second peak remained in a `plateau' for $\sim$ 10 days, before fading to quiescence at 0.21 $\pm$ 0.08 magnitudes day$^{-1}$ over the following 16 -- 28 days (the range being due to the under-sampling of this part of the light curve). 

The recurrence of the second peak to a similar maximum flux within $\sim$ 35 -- 40 days is unique behaviour for an AMXP, although there are some similarities to the persistent source, HETE J1900.1$-$2455, which has been in outburst since its discovery in 2005 \citep{ele08}. Secondary maxima have been noted previously for other LMXBs (e.g. \citealp{chen}) who suggested an increase in accretion rate due to re-processing of X-rays from the initial outburst heating the disc and/or donor star. Alternatively, \citet{tru02} and \citet{tru05} show that tidal instabilities in the disc could cause multiple outburst peaks, but that this phenomenon is not as relevant for systems with large mass ratios such as 00291+51, for which the neutron star is at least a factor of $\sim$ 10 more massive than its companion. A further scenario is that of propagation of cooling/heating waves in a disc that may cause additional maxima in light curves (e.g. \citealp{las}). We note that the smaller scale of AMXPs (shorter orbital periods and smaller discs) and their shorter duty cycles (as opposed to `classical' LMXBs) means that we might expect their discs to both empty and re-fill more quickly, explaining their more rapid duty cycle. However, the previous outburst of 00291+59 was in fact brighter (in both X-rays and optical) but did not display this double-peaked behaviour.

\begin{table*}
\caption{Swift XRT X-ray fluxes.}
\begin{tabular}{lcccccc}
\hline\hline
DATE & MODE & EXPOSURE & FLUX& n$_{H}$& PL$_{ind}$ & PL$_{norm}$ \\
&&(ksec)&(erg cm$^{-2}$ s$^{-1}$)$^{\ast}$&(10$^{22}$ cm$^{-2}$)&&(keV$^{-1}$ cm $^{-2}$ s$^{-1}$) at 1 keV\\
\hline
August 15 & PC	& 1.9 & 3.1 $\pm$ 0.1 $\times$ 10$^{-10}$ & 0.6 $\pm$ 0.1 & 1.7 $\pm$ 0.1 & 8.1 $\pm$ 0.8 $\times$ 10$^{-2}$ \\
August 21 & WT	& 1.2 &	$<$4.7 $\times$ 10$^{-12}$ & -- &	-- & -- \\
August 23 & WT	& 0.4 &	$<$5.5 $\times$ 10$^{-12}$ & -- &	-- & -- \\
August 27 & PC	& 2.2 &	$<$2.8 $\times$ 10$^{-13}$ & -- &	-- & -- \\
August 29 & WT	& 4.9 &	$<$2.1 $\times$ 10$^{-12}$ & -- &	-- & -- \\
September 20 & PC & 1.0 & 1.3 $\pm$ 0.1 $\times$ 10$^{-10}$ & 0.5 $\pm$ 0.1 & 1.6 $\pm$ 0.1 & 2.7 $\pm$ 0.4 $\times$ 10$^{-2}$ \\
October 23 & PC & 2.0 &	$<$5.7 $\times$ 10$^{-14}$ & -- &	-- & -- \\
\hline
 \end{tabular}
 \small

$^{\ast}$Unabsorbed flux in the 2 -- 10keV band. The errors on the fluxes and spectral parameters are 1$\sigma$. The flux upper limits are 3$\sigma$ (using the prescription for low number statistics given by \citealp{geh}) and assume a power law spectrum of photon index 1.5 with n$_H$ = 0.5 $\times$ 10$^{22}$ cm$^{-2}$.
 \normalsize
\end{table*}

\begin{figure}[t]
\resizebox{\hsize}{!}{\includegraphics[width=12cm,angle=270]{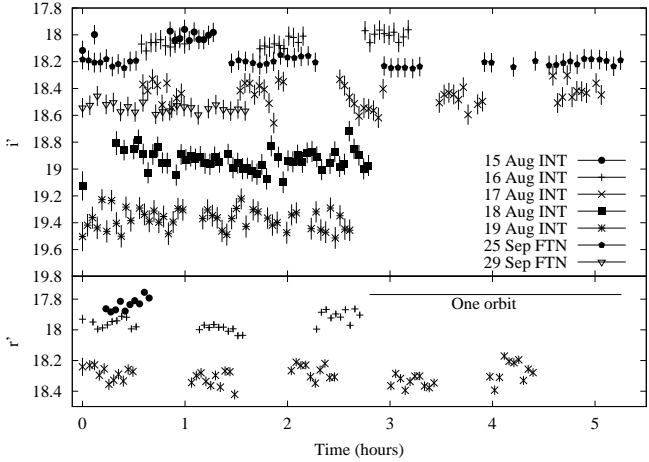}} 
 \caption{\emph{Upper}: 2008 outburst i$^\prime$-band photometry from INT and FTN. \emph{Lower}: 2008 outburst r$^\prime$-band photometry from INT. Error bars shown represent the systematic error of the magnitude; the relative error between data points is smaller.}
 \label{5}
\end{figure}

\begin{table*}
\caption{Variability of 00291+59 and a faint field star.}
\begin{tabular}{lcccccc}
\hline
Date   &\multicolumn{2}{c}{----- Field star -----}&\multicolumn{4}{c}{-------------------------- 00291+59 ------------------------}\\
       &mean mag&1$\sigma$ range      &mean mag&1$\sigma$ range&Significance of variability&Variability level$^1$\\
\hline
15 Aug &i' = 19.127&0.014&i' = 18.016&0.043&3.2 $\sigma$&$43 \pm 14$\\
       &r' = 19.172&0.029&r' = 17.833&0.041&1.4 $\sigma$&$< 130$    \\
16 Aug &i' = 19.121&0.028&i' = 18.043&0.045&1.6 $\sigma$&$< 130$    \\
       &r' = 19.188&0.042&r' = 17.954&0.048&1.2 $\sigma$&$< 170$    \\
17 Aug &i' = 19.094&0.038&i' = 18.447&0.086&2.3 $\sigma$&$86 \pm 38$\\
       &r' = 19.144&0.070&r' = 18.296&0.057&0.8 $\sigma$&$< 270$    \\
18 Aug &i' = 19.109&0.033&i' = 18.935&0.079&2.4 $\sigma$&$79 \pm 33$\\
19 Aug &i' = 19.118&0.032&i' = 19.378&0.077&2.4 $\sigma$&$77 \pm 32$\\
25 Sep &i' = 19.139&0.023&i' = 18.206&0.026&1.0 $\sigma$&$< 110$    \\
29 Sep &i' = 19.140&0.030&i' = 18.543&0.033&1.1 $\sigma$&$< 120$    \\
\hline
\end{tabular}
\small
\\
$^1$The level of variability in milli-mag (when it is detected at the $> 2$ $\sigma$ level), or 3 $\sigma$ upper limit.
\normalsize
\end{table*}

\section{Optical Variability -- orbital, long-term and short-term}

We have monitored 00291+59 since 2007 August and prior to 2008 mid-August, had not detected the source with FTN in i$^\prime$-band, most notably immediately preceding the outburst on August 3 and 9. In 2008 August and September, 00291+59 reached R $\sim$ 18 on two separate occasions. For comparison, its magnitude near the time of discovery in 2004 was R $\sim$ 17. Between the two 2008 peaks, the data are consistent with the source fading to quiescence in i$^\prime$-band (within error bars); the quiescent level in 2004 being I $\sim$ 22.4 \citep{dav,jon}. 

During the August peak, we collected data from the INT in both r$^\prime$ and i$^\prime$-bands (see Fig. 5), over timespans of $\sim$ 3 -- 4 hours (August 18 -- 19) and runs of $\sim$ 30 -- 60 minutes per filter (August 15 -- 17 inclusive). We also observed 00291+59 intensively in i$^\prime$-band around the September peak for $\sim$ 4 hours and $\sim$ 3 hours respectively with FTN. Whilst we note fluctuations of $\sim$ 0.05 magnitudes from one observation to the next (with a cadence of $\sim$ 240 seconds), we saw no evidence on any occasion for any modulation associated with the orbital, or any other period.

This lack of evidence of orbital modulation is likely the result of the light from the accretion disc  swamping much of the lower amplitude orbital variability, and had been noted in the previous outburst \citep{bik,rey,tor08a}. In quiescence, optical observations have twice detected sinusoidal orbital modulation \citep{dav,jon} consistent with emission from an irradiated companion star; this is similar to the example of the AMXP, XTE J1814$-$338 \citep{dav09}, but opposed to SAX J1808.4$-$3658 which displays orbital modulation in both outburst and quiescence \citep{ele}. From the lower panel of Fig. 5, we see that the amplitude of this modulation seems to be greater as the system fades, even though these three datasets are taken on consecutive nights.

In Table 7, we show the level of variability of 00291+59 and a nearby faint field star. Table 7 demonstrates that the variability in each of the INT runs increases as the system fades. We note that we see significant variability in i$^\prime$-band (3$\sigma$) on the first night only (in 2008 August) and that we see less significant variability (2$\sigma$) on 3 of the other 4 i$^\prime$-band observing runs. For the three r$^\prime$-band runs in 2008 August and both i$^\prime$ band runs in 2008 September, we detect variability at less than 2$\sigma$. Since the 2008 outburst, we have continued to monitor 00291+59, but to date (March 2010), have seen no further activity in the system.

\section{Spectral Analysis}

We present a new Keck 1 spectrum (Fig. 6, taken in 2008 August) taken when the source was at R = 21.1 (i.e. when the system was $\sim$ 3 magnitudes fainter than at its maximum, but still 2 magnitudes above quiescence). Since the blue arm suffered from poor signal-to-noise, we display the red arm spectrum only, meaning that we are unable to compare our spectrum with previous detections of H$\beta$, H$\gamma$ or H$\delta$ or the He \small II \normalsize feature at 4686\ \AA\ \citep{tor08a}. We note a prominent double-peaked H$\alpha$ feature at 6563\ \AA\ with a measured equivalent width (EW) of 29.2 $\pm$ 1.7\ \AA\ and FWHM $\sim$ 35\ \AA\ (1600 km s$^{-1}$), indicative of a rotating accretion disc \citep{fra}. The peak-to-peak separation is 18\ \AA\ ($\sim$ 820 km s$^{-1}$) and we see the blue-shifted peak being $\sim$ 48\% stronger than that of the redshifted peak. Double-peaked H$\alpha$ emission has previously been seen in this source \citet{tor08a}, and also in the outbursting AMXP, XTE J1814$-$338 \citep{ste03}. Using our peak-to-peak value, and following the method in Section 10.2 of \citet{tor08a}, we derive an inclination of $\sim$ 35$^{\circ}$, which is consistent with their measurement of 27$^{\circ} \pm 5^{\circ}$.

We do not detect a He\ \small I \normalsize feature at 5875\ \AA\ which\ \citet{tor08a} detected weakly, nor do we see a He \small I \normalsize feature at 6678\ \AA\ which
had been previously noted in one single 300 second spectrum \citep{fil}. At 7065\ \AA\ , we detect a line with low signal-to-noise, which is likely the He \small I \normalsize line as reported in \citet{tor08a} from the spectrum in \citet{rey}. We are also unable to detect the interstellar Na D doublet (5889\ \AA, 5895\ \AA) but do detect a Diffuse Interstellar Band (DIB) at 6284\ \AA\ (EW = 1.47 $\pm$ 0.09\ \AA) as in \citet{tor08a} as well as telluric features at 6864\ \AA\ and 7600\ \AA.

Table 8 summarises the lines measured in all the published optical spectra of this source. From it, we can see that the H$\alpha$ EW is larger at lower optical and X-ray luminosities. We note that this relationship between H$\alpha$ EW and X-ray luminosity is consistent with the previously found trend for neutron star and black hole XRBs \citep{fen09}.

\begin{figure}[t]
\resizebox{\hsize}{!}{\includegraphics[width=12cm,angle=0]{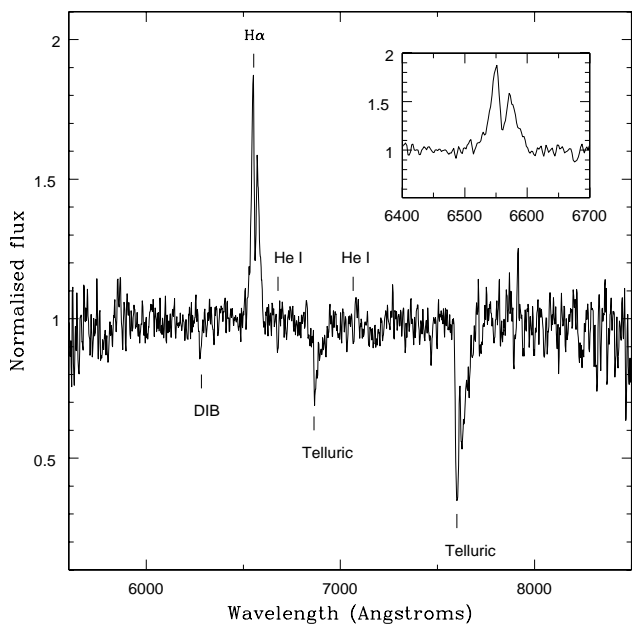}} 
 \caption{Red arm spectrum from Keck 1, DIB denotes Diffuse Interstellar Bands. For clarity, we mark the wavelengths of He \small I \normalsize lines at 6678\ \AA\ and 7065\ \AA\ (the latter being a marginal detection in this spectrum). Insert: zoom in on double-peaked H$\alpha$ emission line profile.}
 \label{6}
\end{figure}

\begin{table}
 \caption{Spectra Equivalent Widths (\AA).}
 \centering
 \begin{tabular}{cccc}
 \hline\hline
Telescope& WHT &Keck I& Keck I\\
Instrument& ISIS& LRIS&LRIS \\
 \hline

Date&2004-12-05&2004-12-12&2008-08-28\\  
 \hline
H$\delta$ 4102&1.2$\pm$0.4&--&--\\
H$\gamma$ 4341&1.7$\pm$0.3&--&--\\
He II 4686&2.9$\pm$0.2& 0.6&--\\
H$\beta$ 4861&2.4$\pm$0.3 & 5.4&--\\
H$\alpha$ 6563&6.5$\pm$0.4 & 9.6 & 29.2$\pm$1.7\\
He I 6678 & --&1.0&--\\
Reference&1&2&3\\ 
\hline
R Mag&17.40&18.61&21.10\\
L$_{\rm X}$(erg s$^{-1})$&8.6 $\times$ 10$^{35}$&1.7 $\times$ 10$^{35}$&$<$2.6 $\times$ 10$^{32}$\\
\hline
 \end{tabular}
 \small
 \newline
1 = \citealp{tor08a}, 2 = \citealp{rey} 3 = this paper. R-band magnitudes and X-ray luminosities are taken within one day of the optical spectra \citep[][this paper]{tor08a}. X-ray luminosities are from RXTE/PCA are based on a distance of 2.8 kpc as adopted in Section 10.3 of \citet{tor08a}.
 \normalsize
\end{table}

\begin{figure*}[t]
 \centering
 \includegraphics[scale=0.35]{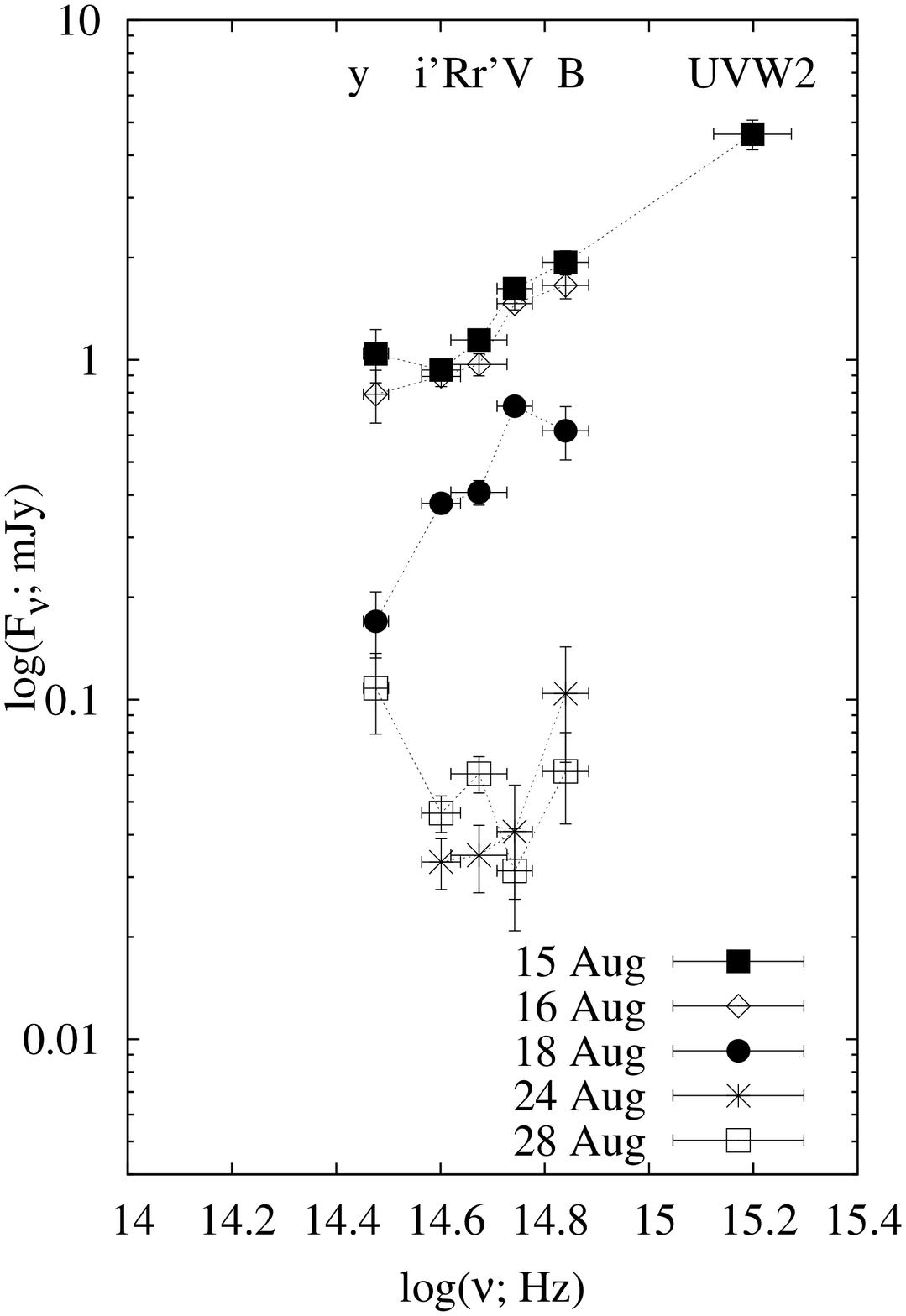}
 \includegraphics[scale=0.35]{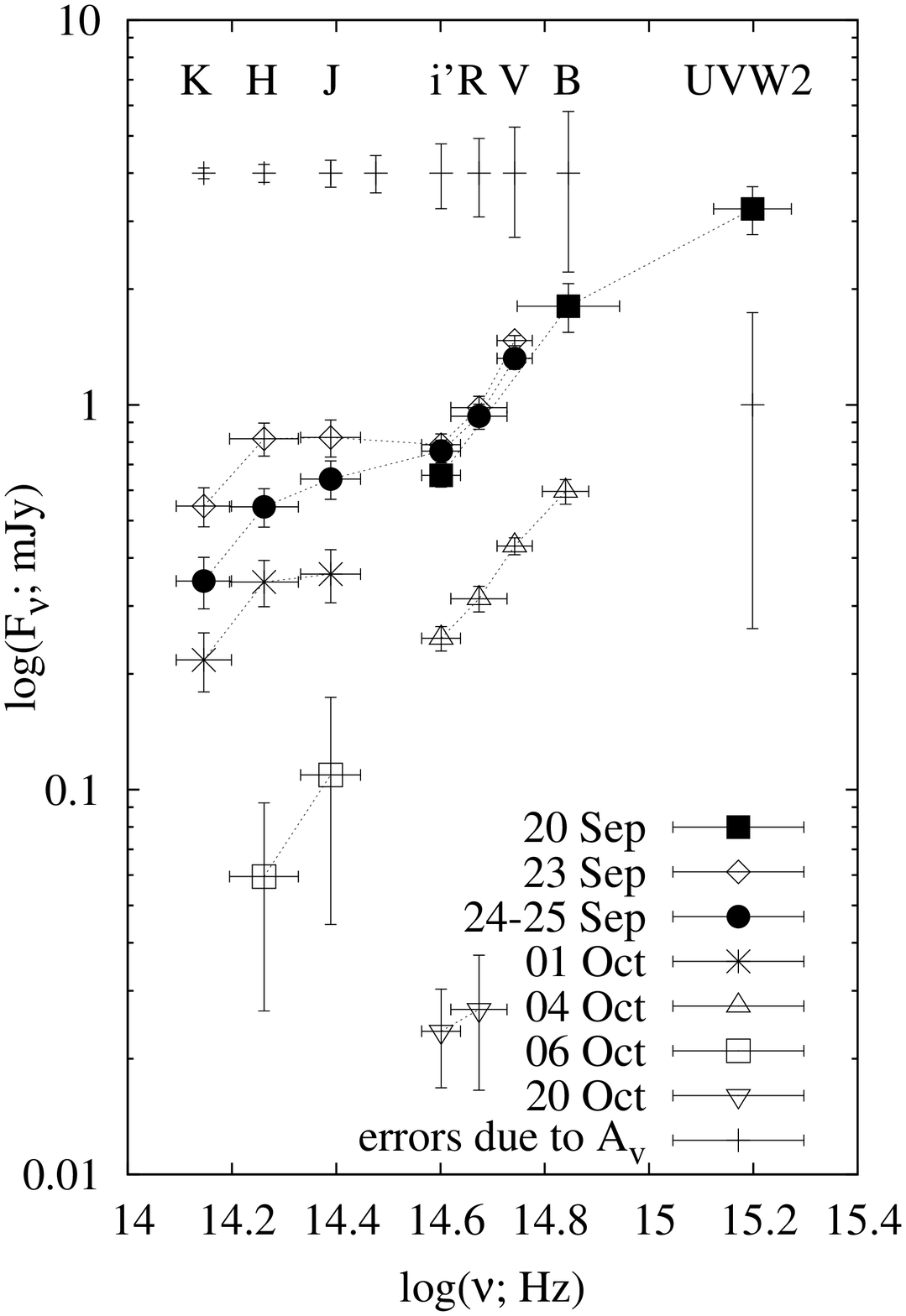}\\
 \includegraphics[scale=0.35]{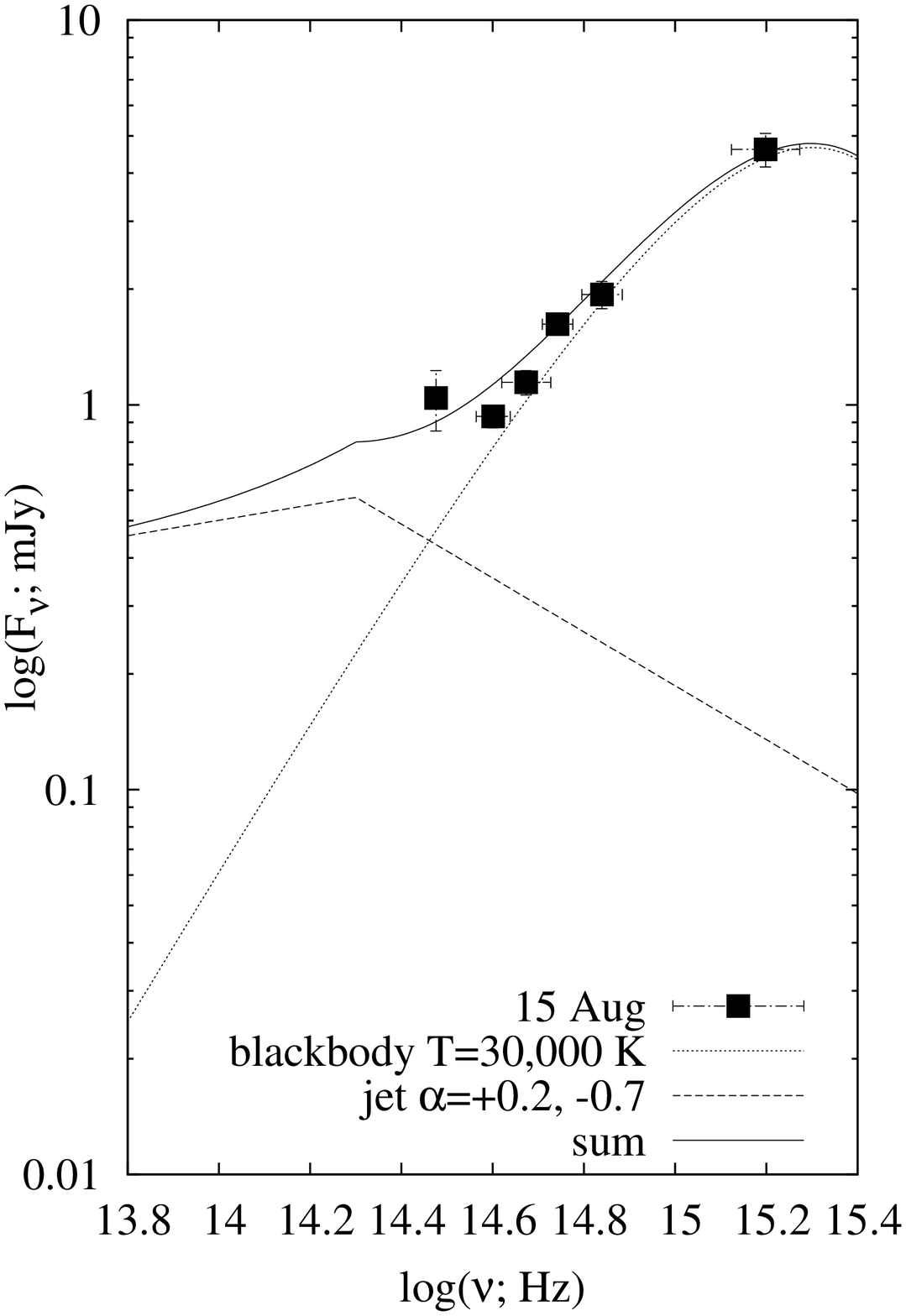}
 \includegraphics[scale=0.35]{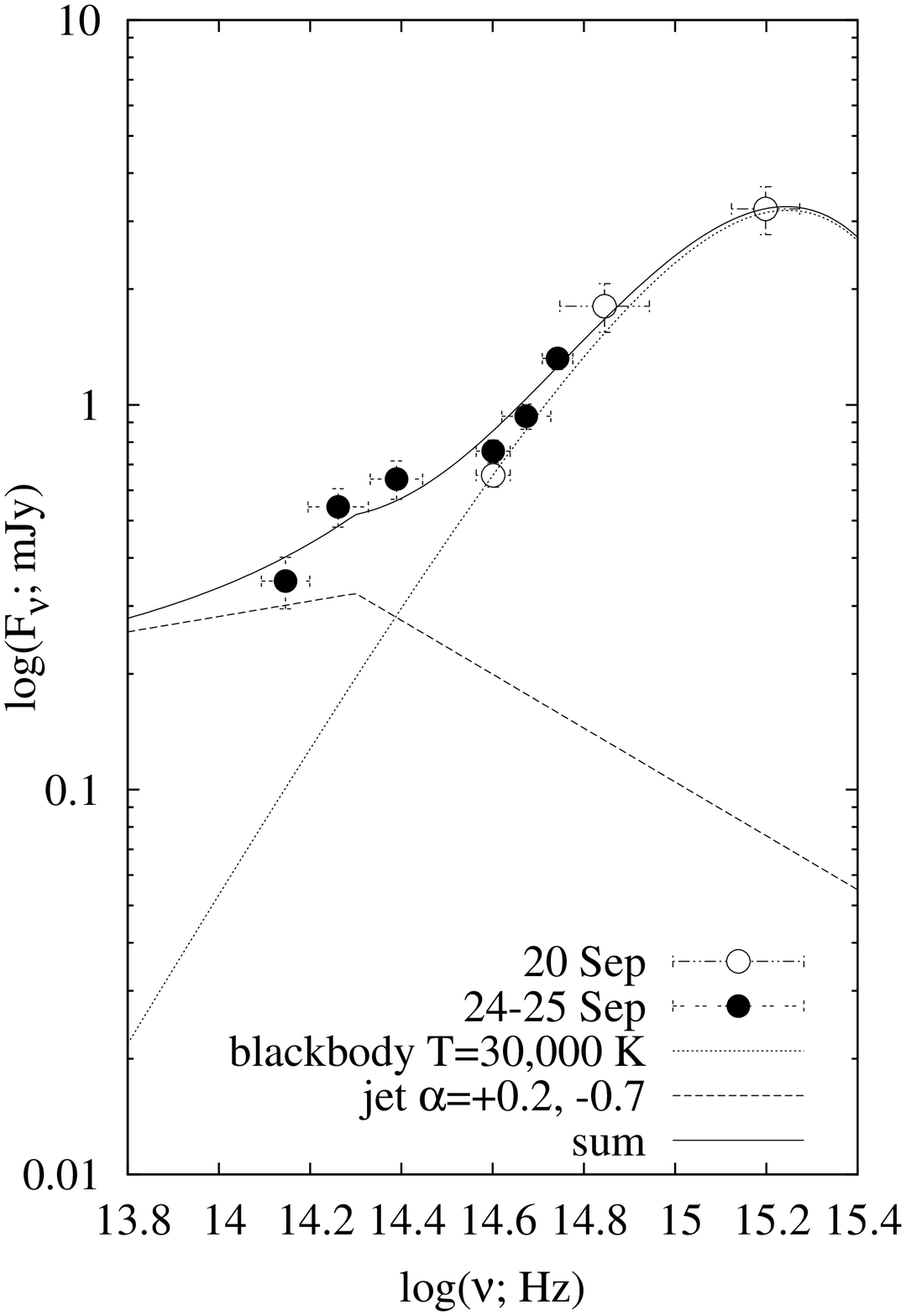}\\
  \caption{\emph{Top row}: UV, optical and NIR fluxes for the 2008 peaks. \emph{Bottom row}: Single black-body plus jet contribution model plotted against fluxes for data at each peak.}
 \label{7}
\end{figure*}

\begin{figure*}[t]
 \centering
 \includegraphics[scale=0.35]{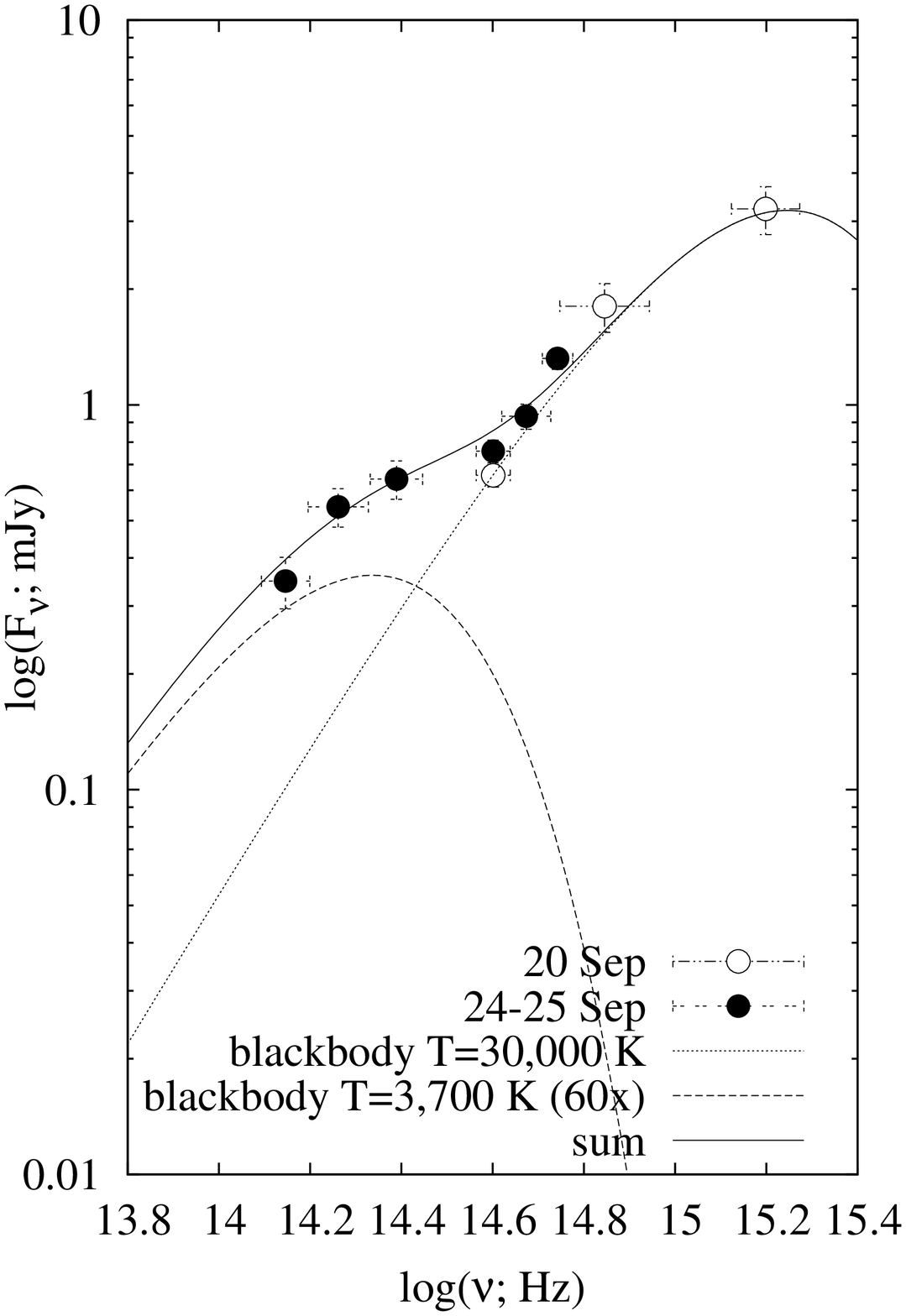}
 \includegraphics[scale=0.35]{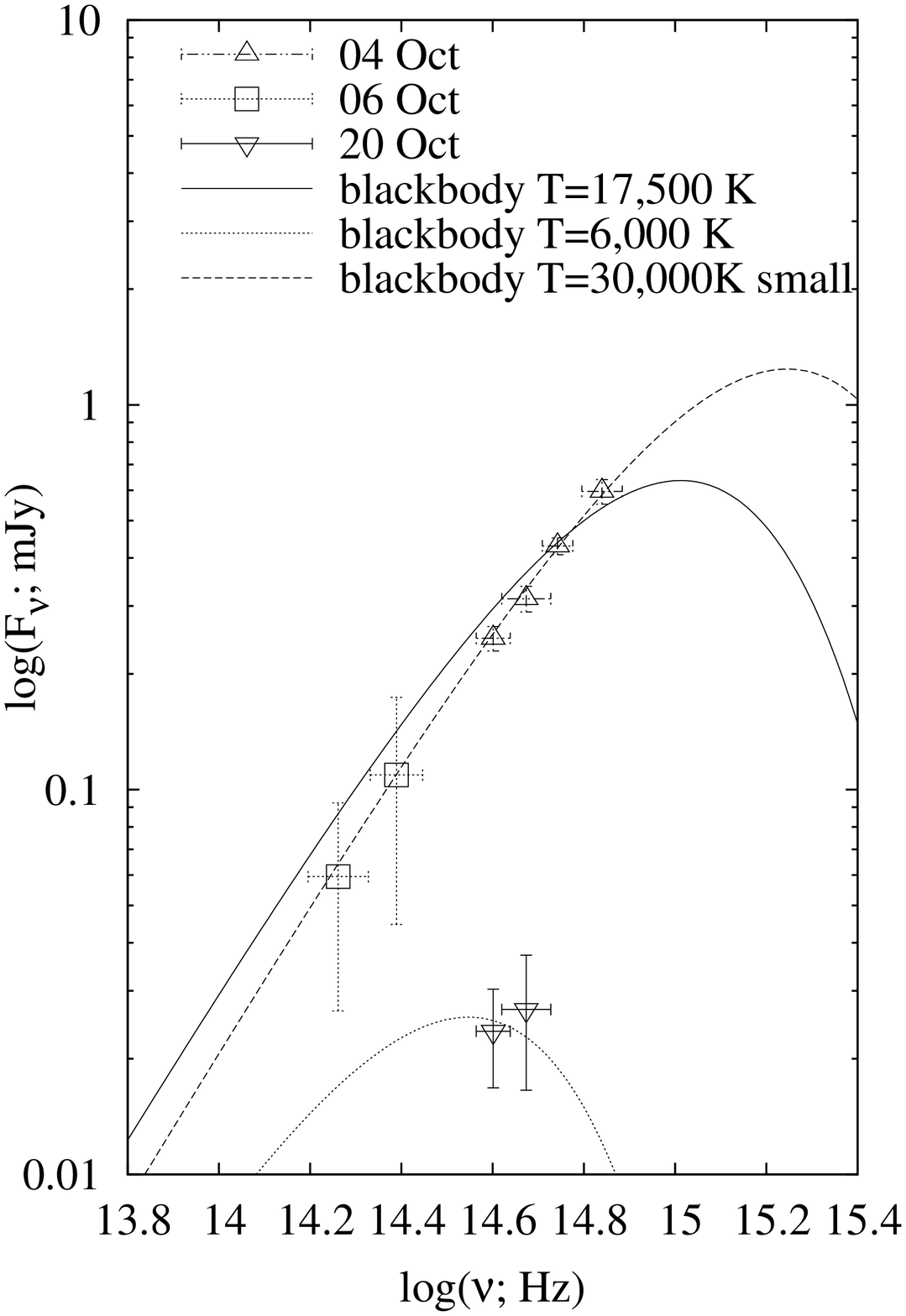}
 \caption{\emph{Left panel}: Model for two black-bodies at second outburst peak. \emph{Right panel}: Model showing the dependence on flux for different temperature disc.}
 \label{8}
 \end{figure*}

\section{Spectral Energy Distributions (SEDs)}

In Fig. 7 we present the de-reddened (see Section 2.1) SEDs compiled from the quasi-simultaneous NIR, optical and UV data from 2008 August (top left panel) and September (top right). These are some of the most complete SEDs ever compiled in this frequency range for neutron star LMXBs (up to eight flux densities measured, spanning one order of magnitude in frequency). We have created SEDs whenever the data were acquired within the same day. The errors in frequency (x-axis) represent the frequency ranges of the filters. For each data point the error in the flux density (y-axis) is derived from the error in the magnitude. We adopt $A_{\rm V} = 2.5$; if the extinction is incorrect by 0.3 mag (see Section 2.1), this propagates into additional errors in intrinsic flux density which are indicated by the separate error bars (crosses) shown across the top of the top right panel, noting that bluer wavelengths are affected more than redder ones. The effect is systematic; if the extinction is $A_{\rm V} = 2.8$ then the intrinsic flux densities were underestimated by the amount indicated by these error bars.

Taking into account these uncertainties, we see that there is generally a blue optical--UV SED ($\alpha > 0$, where $F_{\nu} \propto \nu^{\alpha}$), with a separate NIR excess which dominates the y, J, H, K filters, moreso at higher flux densities. The blue component can be successfully modelled by a single-temperature blackbody of temperature T $\sim 3 \times 10^4$ K in both the August and September peaks (Fig. 7, bottom panels), which is what we might expect from an illuminated or viscously heated accretion disc (e.g. \citealp{hyn}). We are likely sampling a region of the blackbody just redward of the peak, blueward of the Rayleigh-Jeans tail.  However, the errors due to uncertainties in the extinction could significantly change the slope of the SED, and hence the derived blackbody temperature. If the extinction was higher than the measured range of A$_{V}$ = 2.5 $\pm$ 0.3 in \citet{tor08a}, the intrinsic disc spectrum would be bluer than expected from a Rayleigh-Jeans tail, hence confirming the upper limit of  A$_{V}$ $\lesssim$ 2.8.

On the initial fade in August, the SED appears less smooth than for the September decline, in particular a bright y-band flux density on August 28. This may be due to rapid variability, which we do detect to be stronger in the August peak compared to the September peak (see Section 4).

Our October 4 (MJD 54743) observations show that the blackbody is fading, and can either be modelled (Fig. 8, right panel) by a cooler disc of the same area or a hot, smaller disc (implying a change in outer disc size or changes in a disc warp, resulting in a smaller total disc area but at the same temperature; see e.g. \citealp{hyn02}). At lower flux densities later on in the decline of the outburst (October 4,6), the SEDs can be approximated by a blackbody of possibly lower temperature (T $\sim 2-3 \times 10^4$ K). On October 20 (MJD 54759), the source was about one magnitude above its quiescent level in R-band, and at this point, we cannot easily constrain the spectrum, although in Fig.8 (right), we overplot a cooler blackbody of  $\sim 6 \times 10^3$ K.

A NIR excess has been reported for almost all AMXPs in outburst. This excess is most apparent near the outburst peak and is always absent at lower luminosities (see \citealp{rus07} and references therein). It has typically been attributed to synchrotron emission, likely from the jets in the system (e.g. \citealp{wan,gil,kra}). The NIR excess has sometimes been seen quasi-simultaneously with a radio detection; this was the case for 00291+59 in the 2004 outburst \citep{tor08a}.

Here, for the 2008 outburst we also identify a NIR excess above the disc blackbody which seems to disappear later in the outburst when the source fades towards quiescence. However, the excess is not consistent with optically thin synchrotron emission, for which we would expect a red SED with $\alpha \sim -0.7$. Instead, we see an excess which is also blue ($\alpha > 0$), most apparent by a fainter K-band flux density compared to that in H or J. The uncertainties in the extinction cannot explain this fainter K-band flux density. We therefore conclude that the NIR excess in 2008 September is not due to optically thin synchrotron emission.

In the lower panels of Fig. 7 we attempt to model the NIR excess from the outburst peaks in August and September by a simple jet model. Jets in AMXPs usually have a slightly inverted ($\alpha \sim +0.2$) radio-to-optical SED \citep{rus07} from the optically thick (self-absorbed) jet. We take this value of $\alpha$ for the optically thick jet, and $\alpha = -0.7$ for the optically thin jet. If the break between optically thick and optically thin lies around the H-band then the NIR--to--UV SED from September can be explained by this jet (producing the NIR excess) and the blackbody (producing the blue optical--UV emission). Were the break at a lower frequency than H-band, we would expect the K-band to be brighter than observed since it would now be optically thin. Equally, were the break at a higher frequency, then we would expect a smooth increase in flux density (from $\alpha \sim$ 0 to  $\alpha >$ 0) as the optically thick jet spectrum joins the disc spectrum. Instead, the low K-band flux density shows that this cannot be the case \textbf{(although, see below)}. Only a jet break around H-band can produce the NIR excess seen in the SED (on several dates), if we assume the jet is responsible for this NIR excess. The measurement of the break in the jet spectrum is independent of the spectral indices of the jet spectrum's optically thick and thin regimes. Although the data are consistent with this simple model (it is not a fit, but an approximation), this by no means shows that the origin of the NIR excess is the jet. More sophisticated modelling of the broadband data (e.g. \citealp{mai}) are required to confirm whether the NIR excess is consistent or not with a jet origin. Some models predict an `optical bump' in the SED of the synchrotron emission from the jet \citep{mar,per} which could be consistent with these data. 

In the left panel of Fig. 8 we demonstrate that the NIR excess can also be modelled with a second blackbody component. The fainter K-band flux density is more satisfactorily explained in this case, but the blackbody must have a temperature T $\sim 4 \times 10^3$ K and have an area around 60 times that of the accretion disc. The temperature is consistent with (the photosphere of) a low-mass star but the size is not. If the higher temperature blackbody, which dominates the optical, can be explained by the underlying viscously heated disc, the NIR excess may be explained by irradiation of the disc. However, the inferred temperature of this blackbody ($\sim 4 \times 10^3$ K) is far too cool for this to be the case. Alternatively, part of the disc may produce optically thin emission, where the free-free emission has a flatter spectra than the blackbody from the disc. We may expect a circumbinary disc (e.g. \citealp{blu}) surrounding the system to have a temperature and size of this order, but we would not expect a circumbinary disc to be transient; the quiescent NIR flux is $\sim$ two orders of magnitude fainter than this excess. It could be illuminated by the X-ray photons from the central source, but the NIR excess varies more rapidly than the disc, and so seems unlikely. 

\begin{figure*}[t]
 \includegraphics[width=12cm,angle=270]{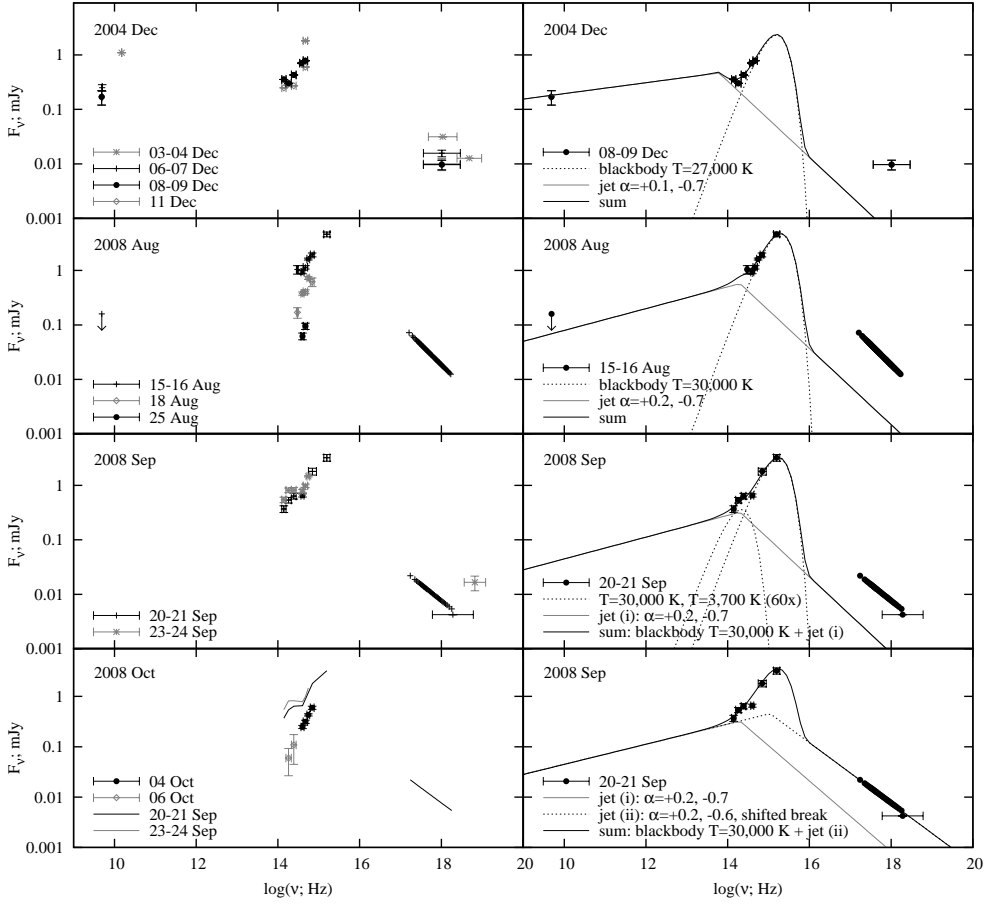}
 \caption{Evolution of the SED of 00291+59. \emph{Left panel}: Broadband SEDs from the 2004 outburst and both 2008 outburst peaks. \emph{Right panel}: Blackbody plus jet contribution shown to approximate data from left panel.}
 \label{9}
\end{figure*}

Quasi-simultaneous, broadband radio--to--X-ray SEDs are presented in Fig. 9. We include data from the 2004 outburst when there were radio detections of the source, because although the data have been published, the broadband SEDs have not. Radio data are taken from \citet{pool,fen04a,rup} and NIR, optical and X-ray data are from \citet{tor08a} and the RXTE ASM.

During the 2004 outburst, a radio counterpart was detected on three dates (peaking at $\sim 1$ mJy) and a blue optical/IR SED with an IR-excess was observed at the same time (Fig. 9, top left panel; see also \citealp{tor08a}). The radio--to--optical SED can be reproduced by the simple blackbody $+$ jet model, adopting $\alpha \sim +0.1$ for the optically thick jet spectrum and a jet break in the mid-IR. The temperature of the blackbody is similar to that found near the peaks of the 2008 outburst, but no UV data were acquired in 2004. Again, this is simply a demonstration that a jet $+$ disc model can satisfactorily reproduce the radio--to--optical SED. We are not claiming this is the broadband spectrum of the jet, or the precise temperature of the disc. For example, it could be that the optically thick jet has a steeper spectral index of $\alpha > +0.1$ and the jet break occurs at lower frequencies. On the other hand, the optically thick jet cannot have a spectral index of $\alpha \sim 0$ if it is to explain the radio data and the IR-excess because the latter component is too bright. It is unlikely that the jet model could explain the X-ray data -- using $\alpha = -0.7$ for the optically thin jet, this component is at least one order of magnitude fainter than the observed X-ray flux. Only if $\alpha \sim -0.4$ for the optically thin jet (which is shallower than typically expected for optically thin synchrotron emission) could this component dominate the X-ray flux. This is similar to the result obtained for the neutron star LMXB 4U 0614+09, for which the synchrotron-emitting jet likely contributes no more than $\sim$ 1\% of the X-ray 2 -- 10 keV flux \citep{migb,mig10}.

In the lower six panels of Fig. 9 we present the radio--to--X-ray SEDs of both declines of the 2008 outburst. The unabsorbed X-ray spectra are provided by the Swift XRT (see Section 2.2.1). The simple blackbody $+$ jet models from close to the outburst peaks are shown in the right hand panels. We again find that if the IR-excess originates in the jet, the synchrotron jet is unlikely to produce the majority of the X-ray emission. If the optically thin jet is extrapolated from the IR-excess to X-ray with $\alpha = -0.7$, the jet could contribute up to $\sim 10$\% of the X-ray flux. In the lower right panel of Fig.9, we show an alternative jet model, jet (ii), in which the optically thick-thin jet break is at a higher frequency ($10^{15}$Hz) and the optically thin spectrum has an index of $\alpha=0.6$. This jet is able to reproduce the X-ray power law (the index being the same as that measured by Swift XRT; see Table 6), but struggles to reproduce the shape of the IR-excess and disc blackbody as well as the first jet model (i) does. We stress that we are unable to make a direct measurement of the synchrotron jet contribution to the X-ray flux; we simply note that with these data, it is possible to reproduce the IR-excess and X-ray powerlaw with a simple jet model, and that more complex modelling (e.g. \citealt{mar}) and probably more complete SEDs are required to make any solid measurement. The radio upper limit near the Aug 2008 outburst peak puts a solid constraint on the radio--to--optical jet spectral index of $\alpha > +0.1$, if the IR-excess originates in the jet, which is consistent with the average value for AMXPs of $\alpha \sim +0.2$ (see discussion above). In the third right panel the second, cool blackbody approximating the IR-excess in 2008 September is also shown.

At first glance it appears the radio jet is intrinsically fainter in 2008 August compared with 2004 December. We investigate this possibility in Fig. 10, in which we plot the radio flux density against the quasi-simultaneous X-ray flux. We find that on the date of the radio observation in 2008 August, the X-ray flux (2 -- 10 keV) is fainter by a factor $\sim 2$ than on the three dates in 2004 on which a radio detection was made. In addition, of the three dates when radio emission was detected, the brightest radio flux was observed on the date of the brightest X-ray flux.

Fig. 10 suggests there may be a relation between radio and X-ray fluxes for 00291+59. The slope of a correlation ($\beta$, where $F_{\rm radio} \propto F_{\rm X}^{\beta}$) is expected to be steeper for neutron star XRBs compared to black hole XRBs (see \citealp{gallo} and references therein). This has been observed with some exceptions (\citealp{miga,tud}).\ \citet{miga} include data of 00291+59 in their analysis but they do not include all the radio detections made in 2004 (or the 2008 upper limit). From the three radio detections, a very steep correlation slope is inferred; $\beta \sim 3.5$. If we neglect the brightest radio detection at 1.1 mJy (see below), the slope is consistent (within errors) with what may be expected (e.g. \citealp{miga}); $\beta = 1.4$ (for NS XRBs) or $\beta = 0.7$ (for BH XRBs). Equally, these slopes are consistent with our non-detection at a lower X-ray flux than those observed in 2004. However, from these data, the correlation slope is poorly constrained.

In black hole XRBs, a bright radio flare is often seen when the source makes a transition to the soft X-ray state, possibly due to a fast, discrete jet ejection slamming into slower jet material launched previously \citep{fen04b}. Although 00291+59 remained in a hard state throughout its 2004 outburst \citep{fen04b}, the 1.1 mJy radio measurement is a factor of three times brighter than the other radio detections (at similar X-ray fluxes). This suggests a discrete jet flare could have been responsible for the brightest detection (as opposed to the supposedly steady, hard state jet). Evidence for ejections in hard states has been found in other XRB transients, e.g. GS 1354$-$64 \citep{bro01}, XTE J1118+480 \citep{bro10}, H1743-322 \citep{jon10}.

The upper limit of 0.16 mJy in 2008 August is consistent with a positive radio--X-ray correlation in 00291+59.

\begin{figure}[h]
\centering
 \includegraphics[height=\columnwidth,angle=270]{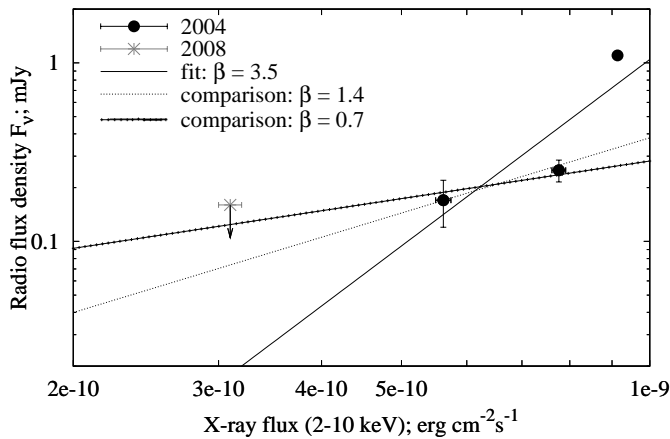}
 \caption{Radio flux density against quasi-simultaneous (within one day) X-ray flux for 2004 and 2008 outbursts. The filled circles represent the 2004 detections \citep{pool,fen04a,rup}; the cross represents the 2008 non-detection (this paper). The solid line \textbf{(without ticks)} shows the power-law fit to all three 2004 detections. For comparison, slopes of $\beta$ = 0.7 and 1.4 as may be expected theoretically are also shown.}
 \label{10}
\end{figure}

\section{Conclusions}

We present a unique multi-wavelength dataset of the 2008 double-peaked outburst of the AMXP, IGR J00291+5934, from radio to X-ray, with coverage of both outburst peaks (2008 August and September). We show that, optically, the September fade is similar in shape to the previous 2004 outburst, however we note a difference between the 2008 August and September peaks in that the former exhibits a more rapid fade, while the latter displays a distinct plateau phase. This plateau, at near maximum luminosity, lasted $\sim$ 10 days before the system faded over a longer timescale than in August. The plateau was also evident in the X-ray light curves; for comparison, the X-ray flux faded by $> 4$ orders of magnitude in $< 10$ days during the first outburst peak in August. From Fig. 4, we see that the morphology of the X-ray fade in both peaks, and in 2004, are very similar. The re-brightening of the system within a few weeks in 2008 is unique within AMXPs, and at odds to this source's previous outburst.

We see short-term variability in the optical of $\sim$ 0.05 magnitudes within the outburst down to a timescale possible of $\sim$ 240 seconds, but no periodicity, on the previously published orbital period. This suggests changes in the accretion flow (likely caused by the irradiation of the disc) dominate over ellipsoidal modulations due to the companion star when the source is in outburst. In addition, the optical variability during the September plateau was of a smaller amplitude than the August peak, at similar flux levels. 

We study the light curve morphology and evolution, presenting the first radio--X-ray Spectral Energy Distributions for this source and the most detailed UV--IR SEDs for any outbursting AMXP. In the optical, the SEDs contain a blue component, which can be fitted by a blackbody, likely from the disc (at maximum luminosity, T $\sim 3 \times 10^4$ K) and a transient near-infrared excess. This excess is consistent with a simple model of a synchrotron jet (as seen in other outbursting AMXPs), however we cannot exclude other potential origins such as the presence of a second, cooler (T $\sim 3.7 \times 10^3$ K) transient blackbody of uncertain origin. We display UV--IR SEDs for 12 dates during the outburst, and find that the IR excess fades more rapidly than the optical disc flux. Our optical spectrum shows the double-peaked H$\alpha$ profile of an accretion disc but we do not clearly see other lines (e.g. He \small I, II\normalsize) that were observed in 2004. 

The lack of a radio detection prevents us drawing many conclusions about the long wavelength part of the SEDs. However, our upper limit is consistent with an optically thick jet (with $\alpha \sim$ +0.2) between radio and NIR, as seen in other AMXPs. Our radio--X-ray SEDs suggest that there is a break in the H-band between an optically thick and optically thin jet. Our simple modelling is also able to account for the radio--optical SED from 2004 which also consists of a blue disc and a NIR-excess \citep{tor08a}. Extending the optically thin jet from infrared--X-ray suggests that the synchrotron jet could account for up to $\sim$ 10$\%$ of the observed X-ray emission, although we note that one synchrotron jet model is able to reproduce the observed hard X-ray power law if the break in the jet spectrum is at a higher frequency than the OIR SED data suggest.

The unusual nature of this double-peaked outburst has implications for the study of AMXPs in particular, and neutron star X-ray transients in general. Not only is this source's double-peaked outburst intriguing, it is difficult to understand what physical processes were different to cause the previous outburst in 2004 to follow a more `traditional' single-peaked morphology. Observing such systems can place tighter constraints on the range of duty cycles of neutron star XRB outbursts. For short period systems such as AMXPs, it allows us to study the evolution of both the outer accretion disc and the inner synchrotron jet, providing data for more accurate modelling of disc-jet coupling in neutron star XRBs. This study illustrates the importance of rapid, regular multi-wavelength monitoring of AMXP outbursts, and of regular optical monitoring of X-ray binaries in quiescence. As was the case with the second outburst peak of IGR J00291+5934, new outbursts can often be detected by such monitoring with 2-metre class telescopes such as the Faulkes Telescopes and YALO/SMARTS (e.g. \citealp{jai,mai07}) before X-ray detection.

\begin{acknowledgements} FL would like to acknowledge support from the Dill Faulkes Educational Trust. DMR acknowledges support from a Netherlands Organization for Scientific Research (NWO) VENI Fellowship. The Faulkes Telescope Project is an educational and research arm of the Las Cumbres Observatory Global Telescope (LCOGT). RXTE/ASM results are provided by the RXTE/ASM teams at MIT and at the RXTE SOF and GOF at NASA's GSFC. The Westerbork Synthesis Radio Telescope is operated by the ASTRON (Netherlands Institute for Radio Astronomy) with support from the Netherlands Foundation for Scientific Research (NWO). PGJ acknowledges support from a VIDI grant from the Netherlands Organisation for Scientific Research. DS acknowledges a STFC Advanced Fellowship. The Peters Automated Infrared Imaging Telescope (PAIRITEL) is operated by the Smithsonian Astrophysical Observatory (SAO) and was made possible by a grant from the Harvard University Milton Fund, the camera loan from the University of Virginia, and the continued support of the SAO and UC Berkeley. We thank the staff and students of Paulet High School (Burton-on-Trent, England), The Kingsley School (Leamington Spa, England), Czacki High School (Warsaw, Poland), St. Brigid's School (Denbigh, Wales) and St David's Catholic College (Cardiff, Wales) for contributing to the Faulkes LMXB Observing Program and Schools' Initiative (FLOPSI) and Alison Tripp for scheduling these observations. We thank the anonymous referee for their comments and swift reply.

\end{acknowledgements}


\begin{thebibliography}{}

 \bibitem[\protect\citeauthoryear{Alpar et al.}{1982}]{alp} Alpar, M.A., Cheng, A.F., Ruderman, M.A., Shaham, J., 1982, Nature, 300, 728

\bibitem[\protect\citeauthoryear{Archibald et al.}{2009}]{arc} Archibald, A.M., Stairs, I.H., Ransom, S.M. et al., 2009, Science, 324, 1411

 \bibitem[\protect\citeauthoryear{Altamirano et al.}{2009}]{alt} Altamirano, D., Patruno, A., Heinke, C.O. et al., 2009, (arXiv:0911.0435)

 \bibitem[\protect\citeauthoryear{Bikmaev et al.}{2005}]{bik} Bikmaev, I., Suleimanov, V., Galeev, A. et al., 2005, ATel, 395

 \bibitem[\protect\citeauthoryear{Blake et al.}{2005}]{bla} Blake, C.H., Bloom, J.S., Starr, D.L., 2005, Nature, 435, 181

  \bibitem[\protect\citeauthoryear{Bloom et al.}{2006}]{blo} Bloom, J.S., Starr, D.L., Blake, C.H., Skrutskie, M.F., Falco, E.E., 2006, in `Astronomical Data Analysis Software and Systems XV', edited by Gabriel, C., Arviset, C., Ponz, D., Solano, E., ASP Conference Series

 \bibitem[\protect\citeauthoryear{Blundell et al.}{2008}]{blu} Blundell, K.M., Bowler M.G., Schmidtobreick, L., 2008, ApJ, 678, L47

 \bibitem[\protect\citeauthoryear{Bozzo et al.}{2010}]{boz} Bozzo, E., Ferrigno, C., Falanga, M. et al., 2010, A\&A, 509, L3

 \bibitem[\protect\citeauthoryear{Brocksopp et al.}{2001}]{bro01} Brocksopp, C., Jonker, P.G., Fender, R.P. et al., 2001, MNRAS, 323, 517

 \bibitem[\protect\citeauthoryear{Brocksopp et al.}{2010}]{bro10} Brocksopp, C., Jonker, P.G., Maitra, D. et al., 2010, MNRAS, in press (arXiv:1001.1965) 

 \bibitem[\protect\citeauthoryear{Burderi et al.}{2007}]{bur} Burderi, L., Di Salvo, T., Lavagetto, G. et al., 2007, ApJ, 657, 961

 \bibitem[\protect\citeauthoryear{Cardelli et al.}{1989}]{car} Cardelli, J.A., Clayton, G.C., Mathis, J.S., 1989, ApJ, 345, 245

 \bibitem[\protect\citeauthoryear{Chakrabarty et al.}{2008}]{chak} Chakrabarty, D., Swank, J.H., Markwardt, C.B., Smith, E., 2008, ATel, 1660

 \bibitem[\protect\citeauthoryear{Chen et al.}{1997}]{chen} Chen, W., Shrader, C.R., Livio, M., 1997, ApJ, 491, 312

 \bibitem[\protect\citeauthoryear{D'Avanzo et al.}{2007}]{dav} D'Avanzo, P.D., Campana, S., Covino, S. et al., 2007, A\&A, 472, 881

 \bibitem[\protect\citeauthoryear{D'Avanzo et al.}{2009}]{dav09} D'Avanzo, P.D., Campana, S., Casares, J. et al., 2009, A\&A, 508, 297

 \bibitem[\protect\citeauthoryear{Eckert et al.}{2004}]{eck} Eckert, D., Walter, R., Kretschmar, P. et al., 2004, ATel, 352

\bibitem[\protect\citeauthoryear{Elebert et al.}{2008}]{ele08} Elebert, P., Callanan, P.J., Fillippenko, A.V., et al., 2008, MNRAS, 383, 1581 

 \bibitem[\protect\citeauthoryear{Elebert et al.}{2009}]{ele} Elebert, P., Reynolds, M.T., Callanan, P.J. et al., 2009, MNRAS, 395, 884

 \bibitem[\protect\citeauthoryear{Fender et al.}{2004a}]{fen04a} Fender, R.P., De Bruyn, G., Pooley, G., Stappers, B., 2004a, ATel, 361

 \bibitem[\protect\citeauthoryear{Fender et al.}{2004b}]{fen04b}  Fender, R.P., Belloni, T.M., Gallo, E., 2004b, MNRAS, 355, 1105

 \bibitem[\protect\citeauthoryear{Fender et al.}{2009}]{fen09} Fender, R.P., Russell, D.M., Knigge, C. et al., 2009, MNRAS, 393, 1608

 \bibitem[\protect\citeauthoryear{Fillipenko et al.}{2004}]{fil} Filippenko, A.V., Foley, R.J., Callanan, P.J., 2004, ATel, 366

 \bibitem[\protect\citeauthoryear{Fox \& Kulkarni}{2004}]{fox} Fox, D.B., Kulkarni, S.R., 2004, ATel, 354

 \bibitem[\protect\citeauthoryear{Frank et al.}{2002}]{fra} Frank, J., King, A., Raine, D., 2002, `Accretion Power in Astrophysics: Third Edition', Cambridge University Press, Cambridge, England

 \bibitem[\protect\citeauthoryear{Gallo et al.}{2003}]{gallo} Gallo, E., Fender, R.P., Pooley, G.G., 2003, MNRAS, 344, 60 

 \bibitem[\protect\citeauthoryear{Galloway et al.}{2005}]{gall} Galloway, D.K., Markwardt, C.B., Morgan, E.H., Chakrabarty, D., Strohmayer, T.E., 2005, ApJ, 622, L45

 \bibitem[\protect\citeauthoryear{Galloway et al.}{2008}]{gall08} Galloway, D.K., Hartman, J.M., Chakrabarty, D.M., Morgan, E.H., Swank, J.H., 2008, ATel, 1786 

 \bibitem[\protect\citeauthoryear{Gehrels}{1986}]{geh} Gehrels, N., 1986, ApJ, 303, 336 

 \bibitem[\protect\citeauthoryear{Giles et al.}{2005}]{gil} Giles, A.B., Greenhill, J.G., Hill, K.M., Sanders, E., 2005, MNRAS, 361, 1180

 \bibitem[\protect\citeauthoryear{Heinke et al.}{2009}]{hei} Heinke, C.O., Altamirano, D., Cohn, H.N., et al., 2009, (arXiv:0911.0444) 

 \bibitem[\protect\citeauthoryear{Hynes et al.}{2002}]{hyn02} Hynes, R.I., Haswell, C.A., Chaty, S., Shrader, C.R., Cui, W., 2002, MNRAS, 331, 169

 \bibitem[\protect\citeauthoryear{Hynes}{2005}]{hyn} Hynes, R.I., 2005, ApJ, 623, 1026

 \bibitem[\protect\citeauthoryear{Jain et al.}{2005}]{jai} Jain, R.K., Bailyn, C.D., Orosz, J.A., McClintock, J.E., Remillard, R.A., 2005, ApJ, 554, 181

  \bibitem[\protect\citeauthoryear{Jonker et al.}{2003}]{jon03} Jonker, P.G., Nelemans, G., Wang, Z. et al., 2003, MNRAS, 344, 201

 \bibitem[\protect\citeauthoryear{Jonker et al.}{2008}]{jon} Jonker, P.G., Torres, M.A.P., Steeghs, D., 2008, ApJ, 680, 615

 \bibitem[\protect\citeauthoryear{Jonker et al.}{2010}]{jon10} Jonker, P.G., Miller-Jones, J.C.A., Homan, J., 2010, MNRAS, 401, 1255

 \bibitem[\protect\citeauthoryear{Krauss et al.}{2005}]{kra} Krauss, M.I., Wang, Z., Dullighan, A. et al., 2005, ApJ, 627, 910

 \bibitem[\protect\citeauthoryear{Landolt}{1992}]{lan} Landolt, A.U., 1992, AJ, 104, 372 

 \bibitem[\protect\citeauthoryear{Lasota}{2001}]{las} Lasota, J.-P., 2001, New AR, 45, 449

 \bibitem[\protect\citeauthoryear{Lewis et al.}{2008a}]{lewa} Lewis, F., Russell, D.M., Fender., R.P. et al., 2008a, in  `Proceedings of the VII Microquasar Workshop: Microquasars and Beyond', 1-5 September 2008, Foca, Izmir, Turkey, Proceedings of Science (arXiv:0811:2336)

 \bibitem[\protect\citeauthoryear{Lewis et al.}{2008b}]{lewb} Lewis, F.,  Linares, M., Russell, D.M., Wijnands, R., Roche, P., 2008b, ATel, 1726

 \bibitem[\protect\citeauthoryear{Linares et al.}{2007}]{lin07} Linares, M., van der Klis, M., Wijnands, R., 2007, ApJ, 660, 595

 \bibitem[\protect\citeauthoryear{Linares et al.}{2008}]{lin} Linares, M., Tudose, V., Migliari, S., 2008, ATel, 1667

\bibitem[\protect\citeauthoryear{Maitra \& Bailyn}{2007}]{mai07}  Maitra D., Bailyn, C.B., 2007, in `Bursts, Pulses and Flickering: wide-field monitoring of the dynamic radio sky', 12-15 June 2007, Kerastari, Tripolis, Greece., Proceedings of Science

 \bibitem[\protect\citeauthoryear{Maitra et al.}{2009}]{mai} Maitra D., Markoff S., Brocksopp C. et al., 2009, MNRAS, 398, 1638

\bibitem[\protect\citeauthoryear{Markoff et al.}{2005}]{mar} Markoff S., Nowak, M.A., Wilms, J., 2005, ApJ, 635, 1203

 \bibitem[\protect\citeauthoryear{Markwardt et al.}{2004a}]{mark04a} Markwardt, C.B., Swank, J.H., Strohmayer, T.E., 2004a, ATel, 353  

 \bibitem[\protect\citeauthoryear{Markwardt et al.}{2004b}]{mark04b} Markwardt, C.B., Galloway, D.K., Chakrabarty, D., Morgan, E.H., Strohmayer, T.E., 2004b, ATel, 360

 \bibitem[\protect\citeauthoryear{Markwardt \& Swank}{2008}]{mark08} Markwardt, C.B., Swank, J.H., 2008, ATel, 1664 

 \bibitem[\protect\citeauthoryear{Markwardt et al.}{2009}]{mar09} Markwardt, C.B., Altamirano, D.,  Swank, J.H. et al., 2009, ATel, 2197

 \bibitem[\protect\citeauthoryear{Marshall et al.}{2008}]{mars08} Marshall, F.E., Markwardt, C.B., Chakrabarty, D., 2008, ATel, 1668

 \bibitem[\protect\citeauthoryear{Migliari \& Fender}{2006}]{miga} Migliari, S., Fender, R.P., 2006, MNRAS, 366, 79

 \bibitem[\protect\citeauthoryear{Migliari et al.}{2006}]{migb} Migliari, S., Tomsick, J.A., Maccarone, T.J. et al., 2006, ApJ,643, L41

 \bibitem[\protect\citeauthoryear{Migliari et al.}{2010}]{mig10} Migliari, S., Tomsick, J.A., Miller-Jones, J.C.A. et al., 2010, ApJ, 710, 117 

 \bibitem[\protect\citeauthoryear{Oke}{1990}]{oke} Oke, J.B., 1990, AJ, 99, 1621

 \bibitem[\protect\citeauthoryear{Patruno et al.}{2009}]{patruno} Patruno, A., Watts, A., Klein-Wolt, M., Wijnands, R., van der Klis, M., 2009, ApJ, 707, 1296

 \bibitem[\protect\citeauthoryear{Pe'er \& Casella}{2009}]{per} Pe'er A., Casella P., 2009, ApJ, 699, 1919

 \bibitem[\protect\citeauthoryear{Pooley}{2004}]{pool} Pooley, G., 2004, ATel, 355

 \bibitem[\protect\citeauthoryear{Radhakrishnan \& Srinivasan}{1982}]{rad} Radhakrishnan, V., Srinivasan, G., 1982, Curr.Sci., 51, 1096

 \bibitem[\protect\citeauthoryear{Remillard}{2004}]{rem} Remillard, R., 2004, ATel, 357

 \bibitem[\protect\citeauthoryear{Reynolds et al.}{2006}]{rey} Reynolds, M.T., Elebert, P., Callanan, P.J. et al., 2006, IAUS, 230, 80

 \bibitem[\protect\citeauthoryear{Roelofs et al.}{2004}]{roe} Roelofs, G., Jonker, P.G., Steeghs, D., Torres, M., Nelemans, G., 2004, ATel, 356

 \bibitem[\protect\citeauthoryear{Rupen et al.}{2004}]{rup} Rupen, M.P., Dhawan, V., Mioduszewski, A.J., 2004, ATel, 364

 \bibitem[\protect\citeauthoryear{Russell et al.}{2007}]{rus07} Russell, D.M., Fender, R.P., Jonker, P.G., 2007, MNRAS, 379, 1108

 \bibitem[\protect\citeauthoryear{Russell et al.}{2008}]{rus} Russell, D.M., Lewis, F., Linares, M., Roche, P., Maitra, D., 2008, ATel, 1666
 
 \bibitem[\protect\citeauthoryear{Sault et al.}{1995}]{sau} Sault R.J., Teuben P.J., Wright M.C.H., 1995, in `Astronomical Data
Analysis Software and Systems IV', edited by Shaw R.A., Payne H.E., Hayes J.J.E., ASP Conference Series

 \bibitem[\protect\citeauthoryear{Shaw et al.}{2005}]{sha} Shaw, S.E., Mowlavi, N., Rodriguez, J. et al., 2005, A\&A, 432, L13

 \bibitem[\protect\citeauthoryear{Steeghs}{2003}]{ste03} Steeghs, D., 2003, IAUC, 8155, 2

 \bibitem[\protect\citeauthoryear{Steeghs et al.}{2004}]{ste} Steeghs, D., Blake, C., Bloom, J.S. et al., 2004, ATel, 363

 \bibitem[\protect\citeauthoryear{Torres et al.}{2008a}]{tor08a} Torres, M.A.P., Jonker, P.G., Steeghs, D. et al., 2008a, ApJ, 672, 1079

 \bibitem[\protect\citeauthoryear{Torres et al.}{2008b}]{tor08b} Torres, M.A.P., Jonker, P.G., Bassa, C., Nelemans, G., Steeghs, D., 2008b, ATel, 1665

 \bibitem[\protect\citeauthoryear{Truss et al.}{2002}]{tru02} Truss, M.R., Wynn, G.A., Murray, J.R., King, A.R., 2002, MNRAS,337, 1329

 \bibitem[\protect\citeauthoryear{Truss}{2005}]{tru05} Truss, M.R., 2005, MNRAS, 356, 1471

 \bibitem[\protect\citeauthoryear{Tudose et al.}{2009}]{tud} Tudose, V., Fender, R.P., Linares, M. et al., 2009, MNRAS, 400, 2111

 \bibitem[\protect\citeauthoryear{Wang et al.}{2001}]{wan} Wang Z., Chakrabarty, D., Roche, P. et al., 2001, ApJ, 563, L61

 \bibitem[\protect\citeauthoryear{Wijnands \& van der Klis}{1998}]{wij98} Wijnands, R., van der Klis, M., 1998, ATel, 17

 \bibitem[\protect\citeauthoryear{Wijnands}{2006}]{wij06}  Wijnands, R., 2006, in  `Trends in Pulsar Research', edited by Lowry, J.A., Nova Science Publishers, New York

 \bibitem[\protect\citeauthoryear{Wijnands et al.}{2008}]{wij} Wijnands, R., Altamirano, D., Soleri, P. et al., 2008, in  `A Decade of Accreting Millisecond X-Ray Pulsars', edited by Wijnands, R., Altamirano, D., Soleri, P., Degenaar, N.,  Rea, N.,  Casella, P.,  Patruno, A. and Linares, M., AIP Conference Proceedings, Springer, Berlin  

 

\end{thebibliography}
\end{document}